\documentclass[useAMS,usenatbib]{mn2e}

\usepackage[totalwidth=500pt,totalheight=680pt]{geometry}

\usepackage{color}
\usepackage{soul}
\usepackage[normalem]{ulem}

\newcommand{\ltsim}{\raisebox{-.5ex}{$\;\stackrel{<}{\sim}\;$}}
\newcommand{\gtsim}{\raisebox{-.5ex}{$\;\stackrel{>}{\sim}\;$}}
\usepackage{graphicx, natbib}
\usepackage{mathrsfs}
\usepackage{subfigure}
\usepackage[font={it,small}]{caption}
\usepackage{listings}
\usepackage{bbding}
\usepackage{amssymb}

\newcommand\aj{{AJ}}%
\newcommand\araa{{ARA\&A}}%
\newcommand\apj{{ApJ}}%
\newcommand\apjl{{ApJ}}%
\newcommand\apjs{{ApJS}}%
\newcommand\aap{{A\&A}}%
\newcommand\aapr{{A\&A~Rev.}}%
\newcommand\aaps{{A\&AS}}%
\newcommand\mnras{{MNRAS}}%
\newcommand\pasp{{PASP}}%
\newcommand{\heii}[1]{{\ensuremath{\mathrm{He}}}\,\textsc{ii}\:#1}
\newcommand{\oiii}[1]{{\ensuremath{\mathrm{O}}}\,\textsc{iii}}
\newcommand{\nii}[1]{{\ensuremath{\mathrm{N}}}\,\textsc{ii}}

\newcommand{\high}[1]{\textcolor{green}{\checkmark}}

\title[AGN in Nearby Low-Mass Galaxies]{The search for active black holes in nearby low-mass galaxies using optical and mid-IR data}
\author[Lia F. Sartori et al.]{Lia F. Sartori$^{1}$\thanks{E-mail:
lia.sartori@phys.ethz.ch}, Kevin Schawinski$^{1}$, Ezequiel Treister$^{2}$, Benny Trakhtenbrot$^{1, \dagger}$, 
\newauthor Michael Koss$^{1, \ddagger}$, Maryam Shirazi$^{1}$, Kyuseok Oh$^{1, 3}$\\
$^{1}$Institute for Astronomy, Department of Physics, ETH Z\"{u}rich, Wolfgang-Pauli-Strasse 27, CH-8093 Z\"{u}rich, Switzerland\\
$^{2}$Universidad de Concepci\'{o}n, Departamento de Astronom\'{i}a, Casilla 160-C, Concepci\'{o}n, Chile\\
$^{3}$Department of Astronomy, Yonsei University, Seoul 120-749, Republic of Korea\\
$^{\dagger}$Zwicky Fellow\\
$^{\ddagger}$Ambizione Fellow}
\begin{document}


\pagerange{\pageref{firstpage}--\pageref{lastpage}} \pubyear{2015}

\maketitle

\label{firstpage}

\begin{abstract}

\textcolor{black} {We investigated AGN activity in low-mass galaxies, an important regime that can shed light onto BH formation and evolution, and their interaction with their host galaxies. We identified 336 AGN candidates from a parent sample of $\sim 48,000$ nearby low-mass galaxies ($M_{\rm \star} \leq 10^{9.5}M_\odot$, $z < 0.1$) in the SDSS. We selected the AGN using the classical BPT diagram, a similar optical emission line diagnostic based on the $\heii{}$$\lambda$4686 line, and mid-IR color cuts. Different criteria select host galaxies with different physical properties such as stellar mass and optical color and only 3 out of 336 sources fulfill all three criteria. This could be in part due to selection biases. The resulting AGN fraction of $\sim 0.7 \%$ is at least one order of magnitude below the one estimated for more massive galaxies.  At optical wavelengths, the $\heii{}$-based AGN selection appears to be more sensitive to AGN hosted in star-forming galaxies than the classical BPT diagram, at least in the low-mass regime. The archival X-ray and radio data available for some of the optically selected AGN candidates seem to confirm their AGN nature, but follow-up observations are needed to confirm the AGN nature of the rest of the sample, especially in the case of mid-IR selection. Our sample will be important for future follow-up studies aiming to understand the relation between BHs and host galaxies in the low-mass regime.} 

\end{abstract}


\section{Introduction}

The formation of Supermassive Black Holes (SMBHs) and their interplay with their host galaxies are fundamental issues in the study of galaxy formation and evolution. Different studies showed that SMBHs with masses $M_{\rm BH} \sim 10^6 - 10^9 M_\odot$ reside in most of the massive galaxies (stellar mass $M_{\rm \star} \gtsim 10^{10} M_\odot$) in the present-day universe, including the Milky Way (e.g. \citealt{Kormendy1995}; \citealt{Schoendel2003}; \citealt{Kormendy2013}). Moreover, SMBHs are thought to grow and evolve in a coordinated manner with the galaxies in which they reside. This idea is supported by the observation of massive galaxies where the SMBH mass is correlated to different host galaxy properties such as stellar mass \citep{Marconi2003}, luminosity (e.g. \citealt{Bemtz2009}; \citealt{Gultekin2009b}), and velocity dispersion (e.g. \citealt{Ferrarese2000}; \citealt{Gebhardt2000}; \citealt{Tremaine2002}; \citealt{Gultekin2009b}) of the bulge.\\

The current observational capabilities do not allow us to directly observe seed BHs and their host galaxies in the early universe. However, dwarf galaxies ($M_{\rm \star} \ltsim 10^{9.5} M_\odot$) are supposed to be similar to the precursors of today's large galaxies and host BHs with masses similar to the first seed BHs \citep{Bellovary2011}. The study of dwarf galaxies hosting massive BHs can thus provide important information about SMBH formation and evolution, as well as their interaction with their host galaxies. 
\textcolor{black} {In addition, the present epoch ($z \sim 0$) BH occupation fraction and mass distribution of low-mass galaxies may help in discriminating between different BH formation scenarios, as discussed by \cite{Volonteri2008a}. In fact, although the numbers are still highly uncertain, different BH formation mechanisms at high redshift predict BH seeds with different properties and evolution, and this will result in a different fraction of galaxies containing SMBHs at later times. The two widely discussed scenarios are heavy seeds from the direct collapse of metal-free gas clouds in the early Universe (e.g. \citealt{Haehnelt1993}; \citealt{Lodato2006}; \citealt{Begelman2006}), and light seeds from the death of massive first generation (Pop III) stars (e.g. \citealt{Bromm1999}, \citealt{Bromm2011}).} \\

One of the most reliable ways to detect a SMBH and to infer its mass is through the analysis of stellar and/or gas dynamics in the host galaxy. However, the gravitational sphere of influence of low-mass BHs in dwarf galaxies is too small to be resolved beyond the Local Group (see \citealt{Reines2013} for more details). \textcolor{black} {This means that BHs with $M_{\rm BH} \sim 10^5 - 10^6 M_\odot$ outside the Local Group can currently be detected only if they are accreting, i.e. if the galaxy hosts an Active Galactic Nucleus (AGN).}

In the last years, two studies have searched for AGN in low-mass galaxies in the Sloan Digital Sky Survey ({\emph{SDSS}}). 
\cite{Reines2013} searched for optical signatures of accreting BHs in a sample of $\sim$25,000 nearby low-mass galaxies ($z < 0.055$, $M_{\rm \star} < 3 \times 10^{9}M_\odot$) selected from the NASA-Sloan Atlas. Applying standard BPT diagram they found an AGN fraction of $\sim$0.13$\%$\footnote{\cite{Reines2013} considered in their analysis both Seyfert (35) and composite (101) objects. The fraction we cite corresponds only to the objects classified as Seyferts.}.
\cite{Moran2014} identified 28 AGN out of a sample of $\sim$7,500 objects in the {\emph{SDSS}} ($0.003 < z < 0.0177$, AGN host galaxies with $M_{\rm \star} < 10^{10}M_\odot$). They derived a lower limit on the AGN fraction of a few percent, but considering a different parent sample than in \cite{Reines2013}.
A different approach was proposed by \cite{Satyapal2014}, who applied mid-IR selection techniques to search for AGN in a sample of bulgeless galaxies. They found that mid-IR selection identifies a different AGN population compared to optical selection, and that the fraction of mid-IR selected AGN is higher at lower masses. 
In a recent study, \cite{Marleau2014} assembled a sample of 313 nearby dwarf galaxies that display mid-IR signature of BH activity. They examined the $M_\star - M_{\rm BH}$ relationship and found that this relation exteds also in the low-mass regime (stellar mass in the range $10^{5}-10^{9}M_\odot$, BH mass in the range $10^{2}-10^{6}M_\odot$).
The literature also offers other studies which attempt to search for low-mass BHs using mid-IR observations (e.g. \citealt{Sat2007,Sat2008,Sat2009}; \citealt{Goulding2010}) or X-ray observations (\citealt{Ghosh2008}; \citealt{Desroches2009a}; \citealt{Gallo2008,Gallo2010}; \citealt{Kam2012}; \citealt{Miller2012}; \citealt{Schramm2013}). Studies aiming to search for AGN with low-mass BHs using optical data were presented by \cite{Greene2004,Greene2007} and \cite{Dong2012}. However, most of these works covered only small volumes with mixed selection functions, and the true BH occupation fraction is poorly understood.\\

In this work, we combine optical emission line and mid-IR data to search for AGN in nearby low-mass galaxies ($z < 0.1$, $M_{\rm \star} < 10^{9.5}M_\odot$), and analyse the properties of their host galaxies. We chose these thresholds for $M_{\rm \star}$ and $z$ because: 1) the difference in the present epoch occupation fraction due to the different formation scenarios is more evident for low-mass galaxies, especially in the mass range $10^{8.5}M_\odot < M_{\rm \star} < 10^{9.5}M_\odot$ (\citealt{Greene2012}, based on models of \citealt{Volonteri2008a}), and 2) dwarf galaxies are faint and can be observed only in the nearby universe.
In this way we measure the fraction of dwarf galaxies containing AGN, placing a lower limit for the BH occupation fraction. Further, assembling a sample of low-mass galaxies hosting AGN is an important step in understanding the relation between BH and host galaxy in the low-mass regime. Because of different biases, the comparison of different selection techniques allows us to find a more complete AGN sample.

The sample selection is described in Section 2. Sections 3 -- 5 describe the analysis techniques performed. The discussion of the results and the summary are found in Sections 6 and 7. Finally, an overview of the AGN candidate samples and of the host galaxy properties is given in the Appendix.

\section{Sample selection}

\subsection{Parent sample of nearby dwarf galaxies}

We assembled a sample of nearby dwarf galaxies in the {\emph{SDSS}} DR7 (\citealt{Abazajian2009}) starting from the {\emph{OSSY}} catalog ({\it{Oh -- Sarzi -- Schawinski -- Yi}}, \citealt{Oh2011}). The {\emph{OSSY}} catalog provides line measurements for the entire spectral atlas from {\emph{SDSS}} DR7 with redshift $z < 0.2$, as well as fitting quality assessment parameters\footnote{The object IDs used in this work refer to the source sequence numbers as they appear in the OSSY database.}.
We selected the objects in the catalog with {\emph{SDSS}} SpecClass = 2 (galaxy) and redshift lower than $z = 0.1$. We then matched the sample to the MPA-JHU catalog \citep{Kauf2003, Brinchmann2004} to obtain the stellar mass, and selected the galaxies with mass lower than $M_{\rm \star} = 10^{9.5}M_\odot$. These masses were derived using fits to photometry and assuming $h = 0.7$. We further excluded $\textcolor{black} {0.17} \%$ of the objects because of unreliable mass estimation. The final sample of nearby dwarf galaxies consists on $\textcolor{black} {48,416}$  objects.

\begin{figure}
\includegraphics[scale=0.4]{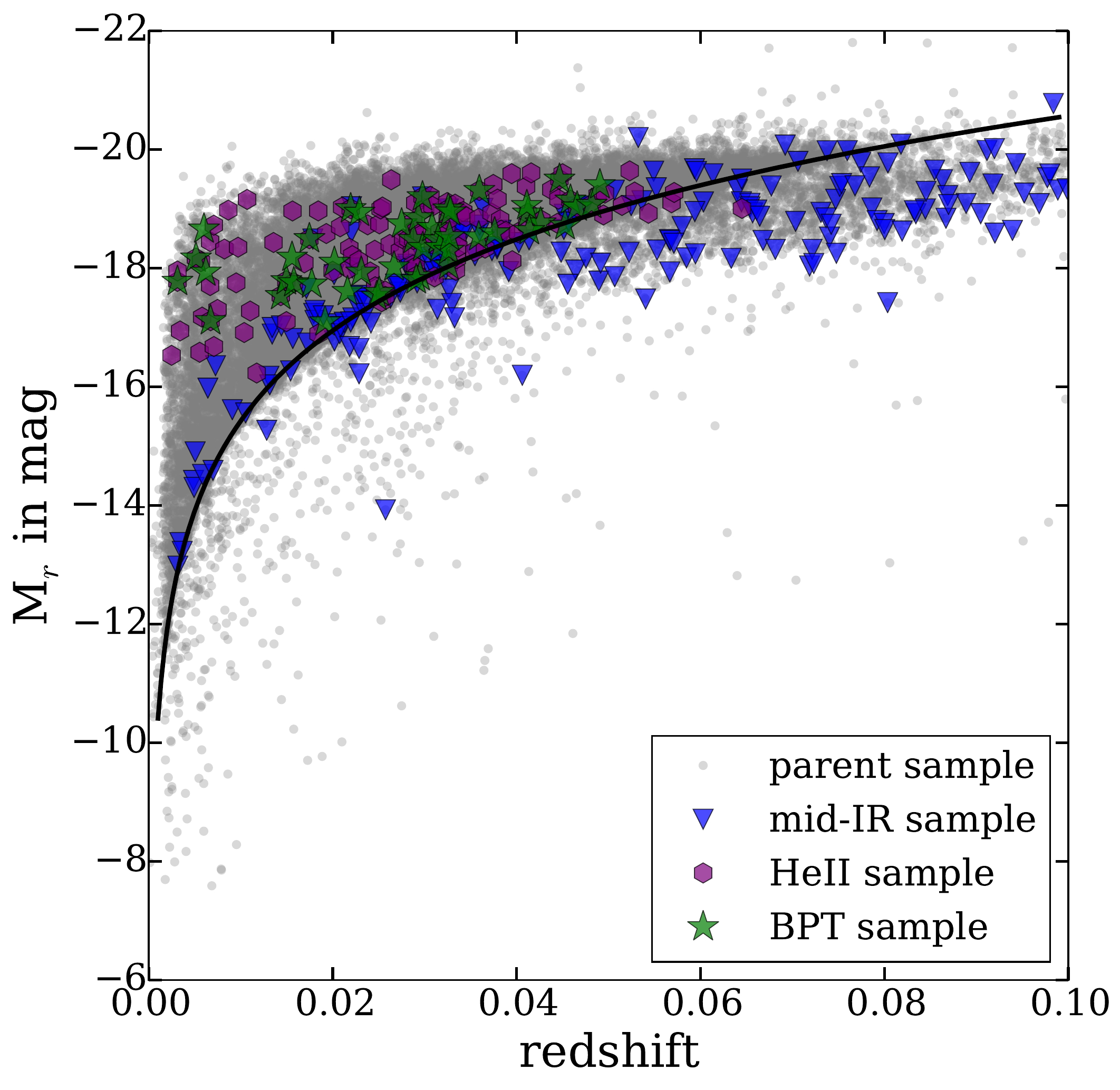}
\caption{Absolute $r$-band magnitude (petrosian) versus redshift for the total sample (gray points) and AGN candidates selected with different selection techniques (see Section \ref{sec:sel}). The black solid line corresponds to the {\emph{SDSS}} spectroscopic apparent magnitude limit, $r = 17.77$ mag. About $\textcolor{black} {15} \%$ of the sample lies below this line and will not be considered for the study of the AGN fraction in dwarf galaxies and of the host galaxy properties.}
\label{fig:mag_red}
\end{figure}

Fig. \ref{fig:mag_red} shows the absolute $r$-band magnitude versus redshift distribution of the nearby dwarf galaxy sample. About $\textcolor{black} {15} \%$ of the sample is fainter than the {\emph{SDSS}} spectroscopic apparent magnitude limit $r = 17.77$ mag (petrosian magnitude). Objects below the spectroscopic limit were often targeted in {\emph{SDSS}} as quasars (e.g. QSO\_CAP, QSO\_SKIRT target), but successively classified as galaxies (SpecClass = 2), so we kept them into the initial parent sample. However, we did not consider objects with $r > 17.77$ mag for the study of the AGN fraction in dwarf galaxies (Section 3) and of the host galaxy properties (Section 4).

\subsection{Selection of AGN candidates}\label{sec:sel}


We searched for AGN in the nearby dwarf galaxy sample by applying three AGN selection techniques, which we discuss in detail in what follows: 
\begin{description}

\item[1)] classical BPT selection (optical emission line diagnostic) based on the separation lines defined by \cite{Kewley2001}, \cite{Kauffmann2003} and \cite{Schawinski2007};

\item[2)] the emission line diagnostic based on $\heii{}$ $\lambda$4686 described by \cite{Shirazi2012} (in the following {\it{Shirazi $\heii{}$ diagram}}); and

\item[3)] the mid-IR color criteria described by \cite{Stern2012} and \cite{Jarrett2011}.

\end{description}

\noindent
In order to define a clean sample of AGN we corrected the obtained sample for fragmentation. In fact, the {\emph{SDSS}} catalog is contaminated by many fragmented pieces of extended massive galaxies, which can be erroneously selected as dwarf galaxies. The number of fragmented galaxies in the parent sample should be low enough to not influence the study of the AGN fraction in dwarf galaxies. However, fragmented galaxies have to be removed from the selected AGN sample. In order to find these objects we performed two tests. First, we compared the stellar mass given in {\emph{MPA-JHU}} with the measurements reported in the {\emph{NASA-Sloan Atlas}}\footnote{http://www.nsatlas.org} ({\emph{NSA}}). Because of the definition of the {\emph{NSA}} catalog  this was possible only for galaxies with redshift $z < 0.055$. We then visually inspected the optical images provided by {\emph{SDSS}}. In this way we excluded $\textcolor{black} {33}$ objects. The list of the AGN candidates is given in Appendix \ref{app:AGN}.

\subsubsection{BPT selection}\label{sec:BPT}

Emission line ratios allow to classify line-emitting galaxies based on their main source of ionization and excitation. The standard diagnostic diagram (\citealt{BPT}, {\it{BPT diagram}}; \citealt{Veilleux1987}; \citealt{Terlevich1991}; \citealt{Kewley2001}; \citealt{Kauffmann2003}; \citealt{Stasinska2006}; \citealt{Schawinski2007}) uses [$\oiii{}$]$\lambda5007{\rm \AA}$/H$\beta$ versus [$\nii{}$]$\lambda6583{\rm \AA}$/H$\alpha$. On this plot, star-forming galaxies lie below the so-called maximum starburst line \citep{Kewley2001} and form a sequence from low metallicity (low [$\nii{}$]/H$\alpha$, high [$\oiii{}$]/H$\beta$) to high metallicity (high [$\nii{}$]/H$\alpha$, low [$\oiii{}$]/H$\beta$). Beyond the maximum starburst line the dominant source of ionization is due to mechanisms other than star formation, such as BH accretion, shocks or evolved stellar populations. The objects residing in this region can be divided into Seyfert galaxies (high ionization emission) and Low Ionization Emission Region galaxies (LINERs, \citealt{Heckman1980}), as proposed by \cite{Schawinski2007} (see also \citealt{Ho2008} and references therein). \textcolor{black} {Because of the different ratio of ionisation emission, in fact, Seyfert galaxies and LINERs show different emission line ratios, and are found in different regions of the diagnostic diagram}. While Seyfert emission is primarily driven by AGN, LINERs can be powered also by fast shocks (\citealt{Heckman1980}; \citealt{Dopita1995}), photoionization by old, metal-rich stellar population (\citealt{Alonso2000}; \citealt{Taniguchi2000}; \citealt{Sarzi2010}), or photoionization by hot stars (\citealt{Filippenko1992}; \citealt{shields1992}; \citealt{Maoz1998}; \citealt{Barth2000}). In our study we consider only Seyfert galaxies as AGN candidates. A further demarcation line, the so-called pure starburst line \citep{Kauffmann2003}, divides pure star-forming galaxies and AGN from composite galaxies, where the ionizing radiation of AGN and star formation is comparable \citep{Panessa2005, Kewley2006, Trouille2011}.

Emission line diagnositics are significantly affected by biases which can cause a significant fraction of AGN in dwarf galaxies to be missed. Low-metallicity AGN have lower [$\nii{}$]/H$\alpha$ ratios and smaller spread in [$\oiii{}$]/H$\beta$ \citep{Kewley2013a, Groves2006, Ludwig2012} and they move to the left part of the BPT diagram, often overlapping with the star-forming region. \cite{Kewley2013a} claimed that AGN are difficult to distinguish from pure star-forming galaxies using the [$\nii{}$]/H$\alpha$ BPT diagram for metallicities lower than $\log(O/H) + 12 \sim 8.4$. Furthermore, BPT selection is not sensitive to blue galaxies with ongoing star formation, which dilutes the signatures of BH accretion. This is an important issue for dwarf galaxies, as they are smaller in the sky and a big fraction of their light falls within the {\emph{SDSS}} fiber (r = 1.5 arcseconds). The detected emission can therefore be dominated by galaxy light from ongoing star formation, dust and gas, so the optical signatures of BH accretion can no longer be detected \citep{Moran2002}.\\

We identified a sample of AGN candidates using the standard BPT diagram, starting from the observed emission line fluxes given in the {\emph{OSSY}} catalog. We decided to perform the BPT selection using the [$\oiii{}$]/H$\beta$ versus [$\nii{}$]/H$\alpha$ diagnostic diagram since [$\nii{}$] is the best indicator for low-metallicity AGN, and dwarf galaxies are expected to have low metallicities (e.g. \citealt{Groves2006}). For this analysis we required all the lines used for the BPT selection to be detected, and amplitude to noise AON $>$ 3 for H$\alpha$, [$\oiii{}$] and [$\nii{}$]. In the cases where only H$\beta$ does not fulfill the requirement AON $>$ 3, we estimated the upper limit of the flux from the standard deviation of the {\emph{SDSS}} spectrum around the line, and assuming the same $\sigma$ value given for the [$\oiii{}$] line ($\sigma_{\rm Balmer}$ in {\emph{OSSY}}). This gives a lower limit for [$\oiii{}$]$\lambda5007{\rm \AA}$/H$\beta$.

\begin{figure*}
\includegraphics[scale=0.25]{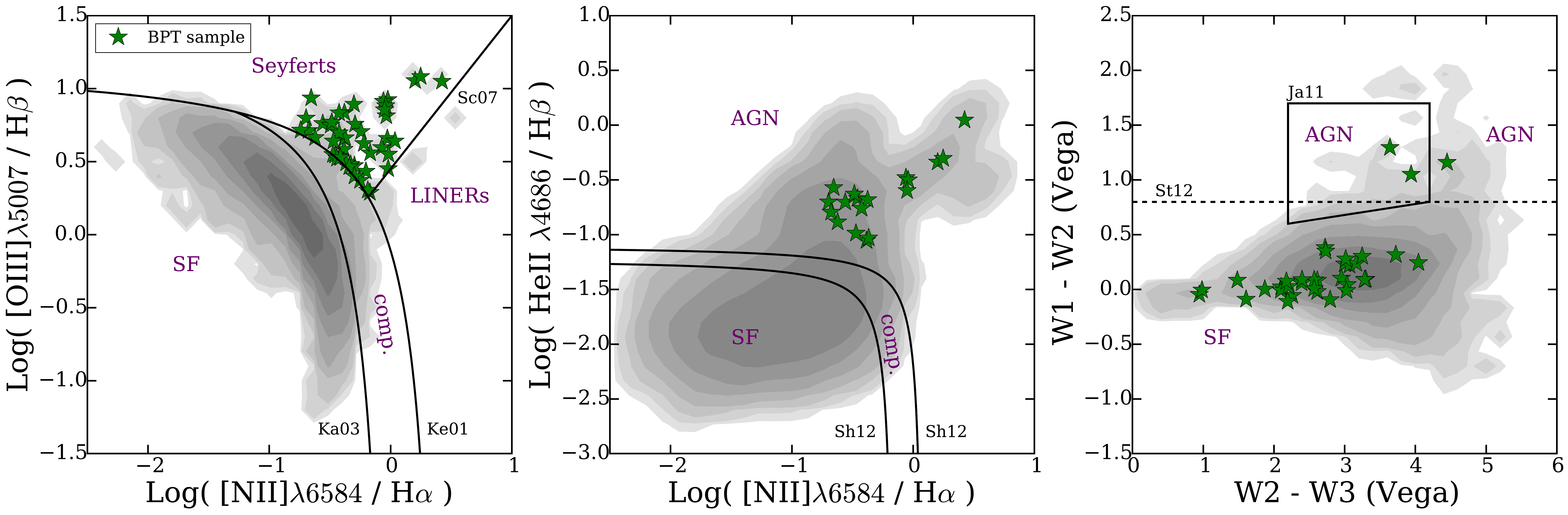}
\caption{Position of the BPT selected AGN candidates on the three selection diagrams applied in this study. The green stars correspond to the BPT selected AGN candidates (48 objects on the BPT diagram), while the gray shaded areas show the position of the objects in the parent sample of nearby dwarf galaxies. On each plot, only the objects which fulfill our AON or SN cuts (see Section \ref{sec:sel} for more details) are plotted. \textit{Left}: BPT diagram based on the separation lines defined by \protect\citet[][\textcolor{black} {maximum starburst line}]{Kewley2001}, \protect\citet[][\textcolor{black} {pure starburst line}]{Kauffmann2003} and \protect\citet[][\textcolor{black} {Seyferts-LINERs separation line}]{Schawinski2007}. \textit{Middle}: $\heii{}$ emission lines diagram as described in \protect\cite{Shirazi2012}. All the BPT selected AGN candidates with well detected $\heii{}$ emission lines are selected as AGN also with this diagram. \textit{Right}: Mid-IR color-color diagram based on {\textit{WISE}} colors. The straight and solid lines show the selection cuts described by \protect\cite{Stern2012} and \protect\cite{Jarrett2011}, respectively. Only 3 BPT selected AGN candidates are selected as AGN also by applying mid-IR color cuts.}
\label{fig:BPT_sel}
\end{figure*}

The results are shown in Fig. \ref{fig:BPT_sel} (left panel). There are $\textcolor{black} {48}$  galaxies in the Seyfert region of the BPT diagram (in the following we refer to these galaxies as {\it{BPT selected AGN candidates}}). Out of the 48 AGN candidates, $\textcolor{black} {40}$  have AON $>$ 3 in all the lines, while $\textcolor{black} {8}$  have upper limits for the H$\beta$ flux\footnote{We computed the upper limits for H$\beta$ as 5 times the noise in the spectrum. This corresponds approximately to a detection with signal-to-noise $S/N = 3$. However the number of AGN selected with H$\beta$ upper limits strongly depends on the upper limits estimation. For example, considering the upper limit as 3 times the noise allow to select $\textcolor{black} {25 }$ additional objects.}. Even with the inclusion of these less reliable candidates, only $\sim \textcolor{black} {0.1}\%$ of the galaxies in the parent sample can be considered as AGN candidate by analysing these specific optical emission line ratios ($\sim \textcolor{black} {0.08}  \%$ if we consider only objects with AON $>$ 3 in each line).


\subsubsection{$\heii{}$ $\lambda$4686}\label{sec:HeII}

As explained above, the classical BPT selection is not very sensitive at low metallicities and for faint AGN, as well as for galaxies with strong ongoing star formation. This can be improved by using lines originating from the mostly neutral interstellar medium, as for example $\heii{}$ $\lambda$4686 (e.g. \citealt{Shirazi2012}). Because of the high ionization energy of He$^+$, $E_{\rm ion} = 54.4$ eV (as a comparison, the ionization potential of O$^{++}$ is 35.5 eV), the $\heii{}$ $\lambda4686{\rm \AA}$ emission line can be produced only in the presence of sources of hard ionization radiation. Therefore, the detection of a luminous $\heii{}$ emission line is a strong indication of AGN activity. 

\begin{figure*}
\includegraphics[scale=0.25]{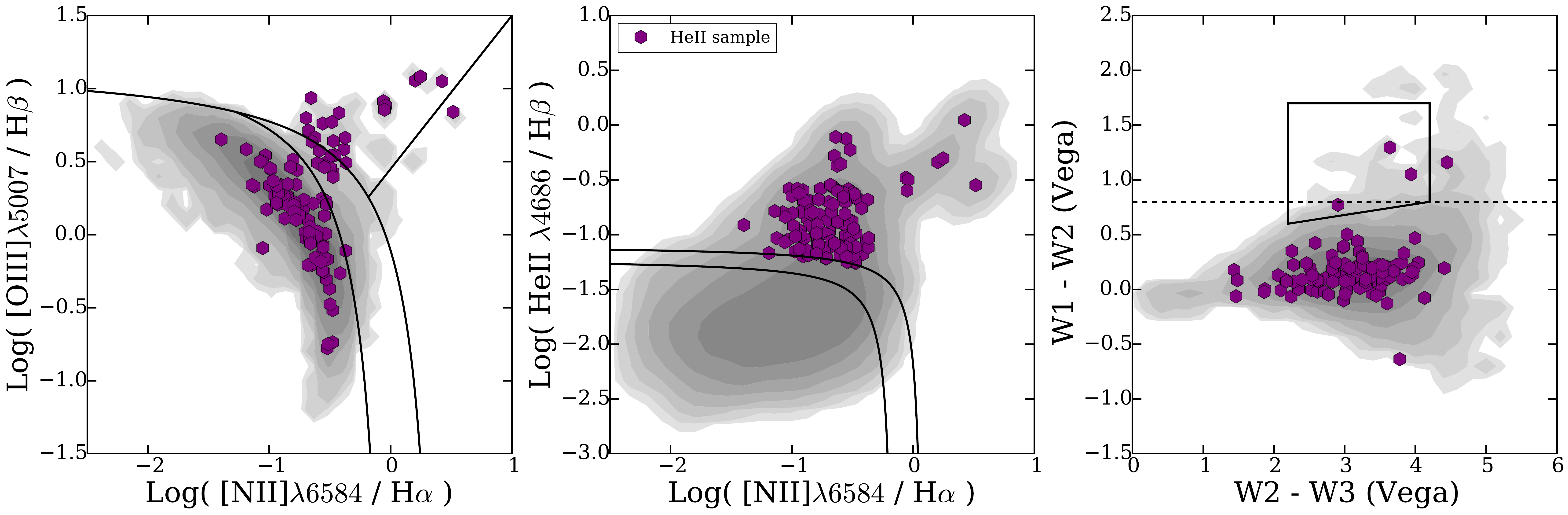}
\caption{Same as Fig. \ref{fig:BPT_sel} but for $\heii{}$ selected AGN candidates (purple hexagons). \textit{Left}: BPT diagram. 18 $\heii{}$ selected AGN candidates are selected as Seyfert galaxies in the BPT diagram. \textit{Middle}: Shirazi $\heii{}$ diagram. All the objects found above the pure starburst line described in \protect\cite{Shirazi2012} are selected as AGN (121 objects). \textit{Right}: Mid-IR color-color diagram. Only 4 of the $\heii{}$ selected AGN candidates are selected as AGN also by applying mid-IR selection.}
\label{fig:HeII_sel}
\end{figure*}

We plotted the $\heii{}$ $\lambda4686{\rm \AA}$/H$\beta$ line ratio versus [$\nii{}$]$\lambda6583{\rm \AA}$/H$\alpha$ diagram described in \cite{Shirazi2012}, and used it as an alternative emission lines diagnostic. Similar to the standard BPT diagram, star-forming galaxies are expected to be below a maximum starburst line, while AGN\footnote{No distinction is made here between Seyferts and LINERs.} are found above this line. We required an amplitude to noise AON $>$ 3 for HeII, [$\nii{}$], H$\beta$ and H$\alpha$. With this technique selected $\textcolor{black} {121  }$ AGN candiates, as shown in Fig. \ref{fig:HeII_sel} (middle panel), which correspond to $\sim \textcolor{black} {0.25}\%$ of the parent sample. In the following we refer to these galaxies as {\it{$\heii{}$ selected AGN candidates}}. 

\begin{figure*}
\includegraphics[scale=0.25]{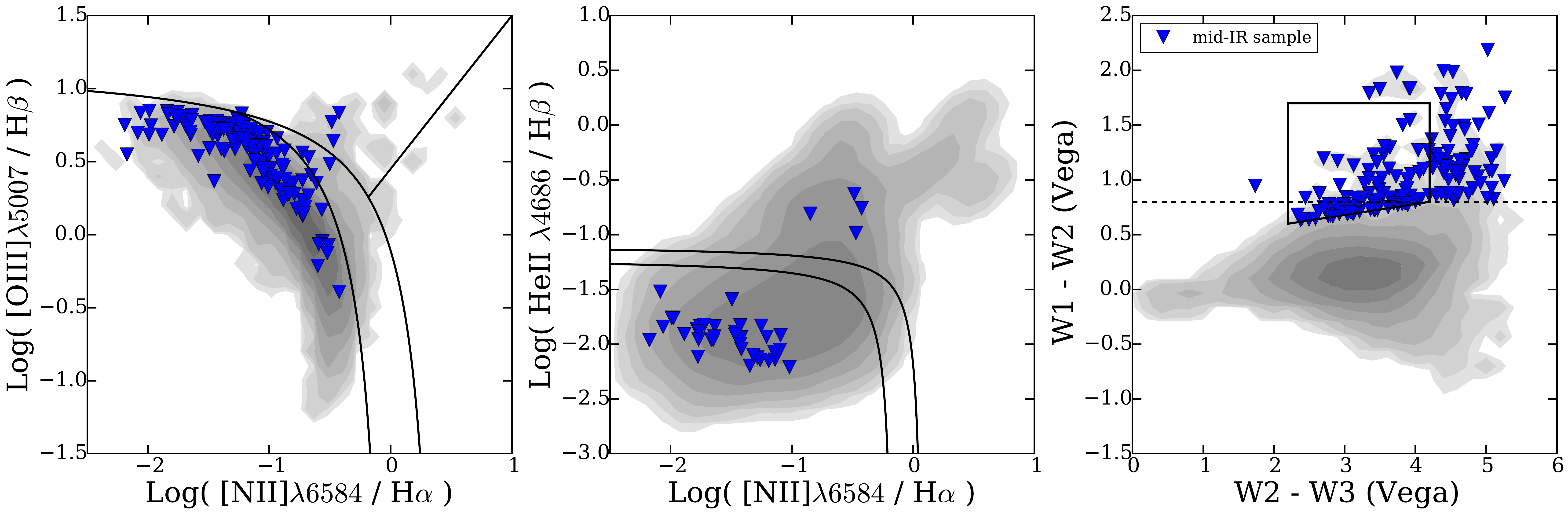}
\caption{Same as Fig. \ref{fig:BPT_sel} but for mid-IR selected AGN candidates (blue triangles). \textit{Left}: BPT diagram. Most of the mid-IR selected AGN candidates are located in the upper part of the star-forming region, near to the demarcation line. \textit{Middle}: Shirazi $\heii{}$ diagram. Only $\textcolor{black} {4}$ mid-IR selected AGN candidates reside on the AGN region of the Shirazi $\heii{}$ diagram. \textit{Right}: Mid-IR color-color diagram. All the objects fulfilling Stern and/or Jarrett criteria are selected as AGN candidates (189 objects).}
\label{fig:color_sel}
\end{figure*}

\subsubsection{Mid-IR color selection}\label{sec:color}

Mid-IR selection relies on the fact that AGN show redder mid-IR colors compared to star-forming galaxies, especially at low redshift \citep{Stern2012,Assef2012}. The hot dust surrounding the AGN produces a strong mid-IR continuum which can be approximated as a power-law dominating in the redder part of the spectrum. On the other hand, the stellar emission is dominated by a composite black body curve which peaks at rest frame $\sim$1.6$\mu$m. An advantage of mid-IR selection compared to selections based on other wavelength ranges is that mid-IR emission is independent of the viewing angle. This allows to select both unobscured (Type 1) and obscured (Type 2) AGN.

Several mid-IR diagnostic diagrams have been proposed and employed in the literature (e.g. \citealt{Lacy2004,Assef2010,Stern2005,Stern2012,Jarrett2011}). In this work we concentrate on cuts which purely rely on the {\it WISE} bands at 3.4, 4.6, 12 and 22 $\mu$m (\emph{W1, W2, W3} and \emph{W4} bands, respectively; \citealt{Wright2010}). \cite{Stern2012} proposed a cut based on the \emph{W1} and \emph{W2} bands, the two bluest, most sensitive channels. Using IRAC-selected AGN as their input sample \citep{Stern2005} they empirically defined a simple color criterion: 

\[[W1 - W1] \ge 0.8.\]

\noindent
For this criterion they estimated a reliability of $95 \%$.
A more restrictive criterion was defined by 
\cite{Jarrett2011}, which also considers the \emph{W3} band:

\[ 2.2 < [W2 - W3] < 4.2\]
\[ (0.1 \times [W2 - W3] + 0.38) < [W1 - W2] < 1.7.\]

Even though mid-IR selection is sensitive to both obscured and unobscured AGN, none of the proposed cuts can provide a complete sample of AGN. For less powerful AGN the dilution by the host galaxy may cause a bluer ($W1-W2$) color, so that these AGN are no longer distinguishable from normal galaxies.\\  

We performed mid-IR color selection following the color cuts described by \cite{Stern2012} and \cite{Jarrett2011}. \textcolor{black} {For this we used the WISE magnitudes measured with profile-fitting photometry as listed in the WISE All-Sky Data Release (\citealt{WISE})}. For the Stern selection we required the objects to be detected in the {\emph{W1}} and {\emph{W2}} bands (in the {\it WISE} catalog this means S/N $>$ 2). For the Jarrett cut we additionally required the detection in the \emph{W3} band. The results are shown in Fig. \ref{fig:color_sel} (right panel). Combining the objects selected by the Stern and Jarrett criteria we found $\textcolor{black} {189}$ mid-IR candidates. Therefore, $\sim $\textcolor{black} {0.4}$ \%$ of the parent sample exhibits mid-IR colors consistent with AGN emission. In the following analysis we will refer to the objects selected with Stern and/or Jarrett criteria as {{\it mid-IR selected AGN candidates}}.

\textcolor{black} {We decided to use the magnitudes measured with profile-fitting photometry in order to be consistent with \cite{Stern2012}, but it is important to notice that these magnitudes could be wrong for some extended nearby galaxy. Among the 189 mid-IR selected AGN candidates only 1 is flagged as possibly extended. However, we examined the WISE images by eye and the extension in {\emph{W1}}, {\emph{W2}} and {\emph{W3}} is minimal, so that it will not affect the result significantly.}

The {\emph{SDSS}} spectra of some mid-IR selected AGN required further fitting. For these objects we performed our own fits and computed the observed flux for $\heii{}$ $\lambda$4686, H$\beta$, [$\oiii{}$]$\lambda$4959, [$\oiii{}$]$\lambda$5007, H$\alpha$, [$\nii{}$]$\lambda$6548 and [$\nii{}$]$\lambda$6583. We fitted the lines with a Gaussian profile, using a fixed ratio between the two [$\oiii{}$] lines and between the two [$\nii{}$] lines.

\subsubsection{Comparison between the selection techniques}\label{sec:comp}

The different selection techniques revealed different samples of AGN candidates. In Fig. \ref{fig:BPT_sel}, \ref{fig:HeII_sel} and \ref{fig:color_sel} we overplotted the AGN candidates selected with each selection technique on the other diagnostic diagrams. 

Among $\textcolor{black} {48}$  BPT selected AGN candidates (Fig. \ref{fig:BPT_sel}), only $\textcolor{black} {3}$  present mid-IR colors attributable to AGN activity. On the other hand, all the BPT selected AGN candidates with well detected $\heii{}$ emission lines are selected as AGN also by the $\heii{}$ criterium.

In a similar manner, only $\textcolor{black} {4}$ of the $\heii{}$ selected AGN candidates (Fig. \ref{fig:HeII_sel}) are selected as AGN also by applying mid-IR selection. On the BPT diagram, 18 $\heii{}$ selected AGN candidates are found in the Seyfert region. Here it is important to notice that the requirement AON $>$ 3 for the $\heii{}$  line is fulfilled only for $\textcolor{black} {1.7} \%$ of the objects in the parent sample of nearby dwarf galaxies. On the other hand, $\textcolor{black} {37.5} \%$ of the BPT selected AGN candidates and $\textcolor{black} {21.9} \%$ of the mid-IR selected AGN candidates show strong $\heii{}$ emission lines.

Consistently with what described above, most of the mid-IR selected AGN candidates (Fig. \ref{fig:color_sel}) are located in the upper part of the star-forming region, near the demarcation line. Only $\textcolor{black} {3}$ mid-IR selected AGN candidates reside on the AGN region of the BPT diagram, and 4 in the AGN region of the Shirazi $\heii{}$ diagram.

Possible explanations of why AGN candidates found with one selection technique are not selected by the other techniques are discussed in Section \ref{sec:disc}.

\section{AGN fraction in dwarf galaxies}

\begin{figure*}
\includegraphics[scale=0.4]{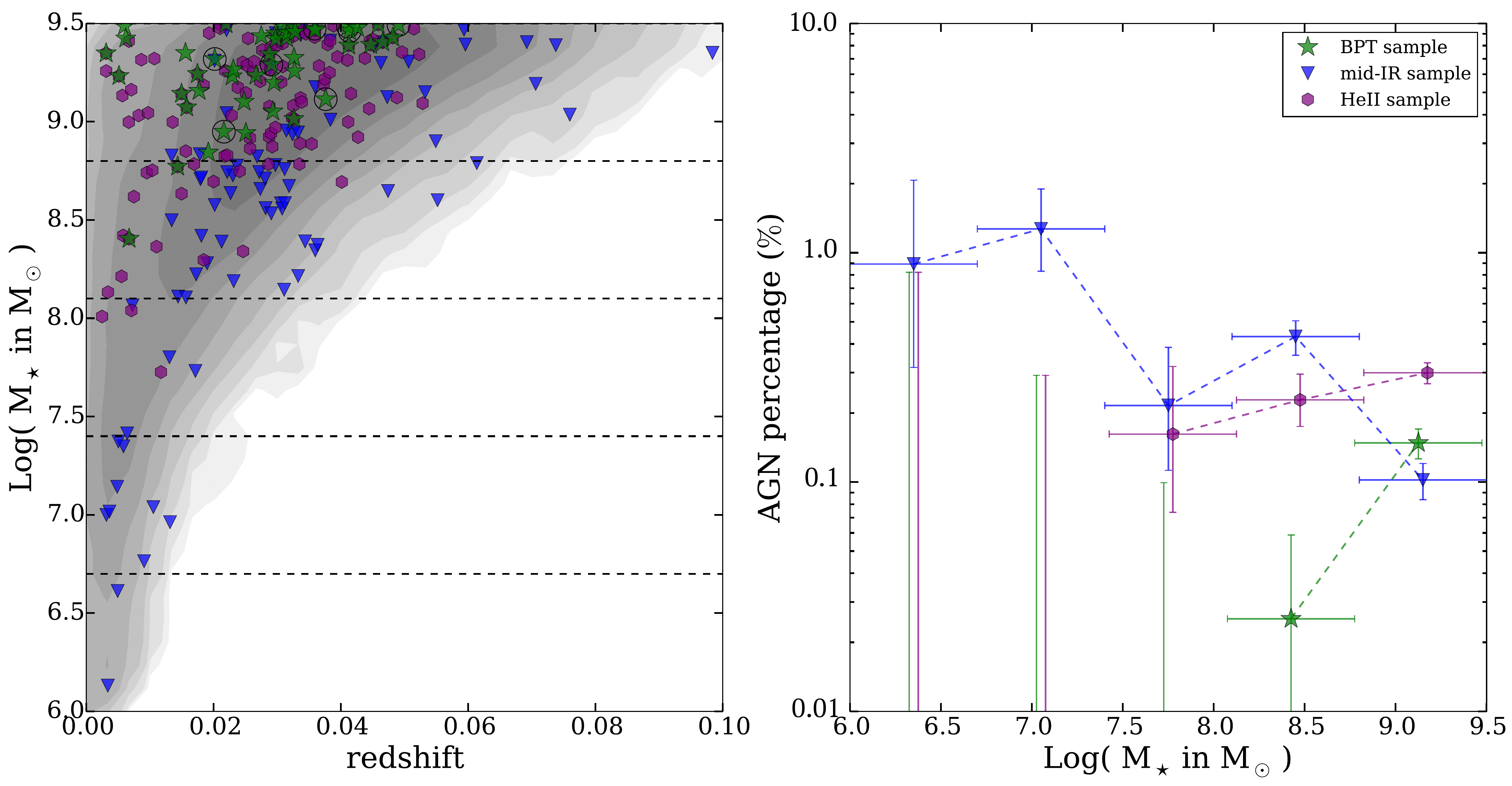}
\caption{\textit{Left}: Mass versus redshift distribution for the parent sample (gray shaded area) and the AGN candidates. Green stars refer to AGN candidates selected with BPT diagram \textcolor{black}{(the circled stars show the AGN candidates with upper limits for the H$\beta$ flux)}, purple hexagons to $\heii{}$ selected AGN candidates, and blue triangles to the mid-IR candidates. Only objects with $r$-band magnitude $r < 17.77$ are considered. The dashed lines divide the sample in five mass bins. \textit{Right}: AGN percentage in the five mass bins defined in the left panel. For each bin we computed the uncertainties of the AGN fraction using binomial statistic (see text for deteails). The three AGN samples are considered separately.}
\label{fig:frac_mass}
\end{figure*}


As showed in Section \ref{sec:sel}, the fraction of AGN found in the parent sample depends on the considered selection method. About $\textcolor{black} {0.3} \%$ of the sample exhibits optical narrow-lines signatures of accreting black holes: $\textcolor{black} {0.1} \%$ is selected through BPT diagram while $\textcolor{black} {0.25} \%$ through the Shirazi $\heii{}$ diagram. On the other hand, the fraction of galaxies with mid-IR colors attributable to AGN is $\sim \textcolor{black} {0.4} \%$. The comparison between the optical and mid-IR selected AGN samples shows almost no overlap: among $\textcolor{black} {336}$ AGN candidates, only $\textcolor{black} {3}$ are selected by all the three selection techniques.

The difference observed between the three samples could be explained from the different biases suffered by the selection criteria (see Section \ref{sec:disc} for a discussion). Moreover, a problem in determining the real AGN fraction is the selection function of the sample, which in our case is a consequence of the {\emph{SDSS}} spectroscopic apparent magnitude limit $r = 17.77$ (petrosian magnitude). In order to perform a reliable demographic analysis we required a magnitude limited sample. Therefore we excluded from the sample all the objects with $r$-band magnitude r $>$ 17.77 mag (see Fig. \ref{fig:mag_red}). In this way we excluded $\textcolor{black} {1}$ BPT selected AGN candidate (out of $\textcolor{black} {48}$), $\textcolor{black} {9}$ $\heii{}$ selected AGN candidates (out of $\textcolor{black} {121}$) and $\textcolor{black} {112}$ mid-IR selected candidates (out of $\textcolor{black} {189}$). The high number of mid-IR selected candidates with r $>$ 17.77 mag can be explained by the fact that the mid-IR color-color diagram is more sensitive to blue objects (optical colors, see Section \ref{sec:host}). These objects were observed by {\emph{SDSS}} because they were erroneously targeted as quasars.

Fig. \ref{fig:frac_mass} shows the AGN percentage in different mass bins. Green stars refer to AGN candidates selected with BPT diagram, purple hexagons to $\heii{}$ selected AGN candidates, and blue triangles to the mid-IR selected AGN candidates. For each bin we computed the uncertainties of the AGN fraction using binomial statistics. In the cases where less than 20 AGN candidates are in the bin, we computed the uncertainty on the AGN number N$_{\rm AGN}$ using the values reported by \cite{Gehrels1986}. For N$_{\rm AGN} \ge $ 20, we applied the formula $\sqrt{f (1-f) N_{\rm tot}}$, where $f$ is the AGN fraction and $N_{\rm tot}$ the total number of objects in the considered bin. The AGN percentage is $< 2 \%$ in each considered bin. For mid-IR selected AGN candidates there is an indication of degreasing fraction with increasing mass. On the other side, BPT and $\heii{}$ selected AGN candidates are found only at higher masses.

\section{Host Galaxy properties}\label{sec:host}

\begin{figure*}

\begin{minipage}{\textwidth}
\centering
\includegraphics[scale=0.3]{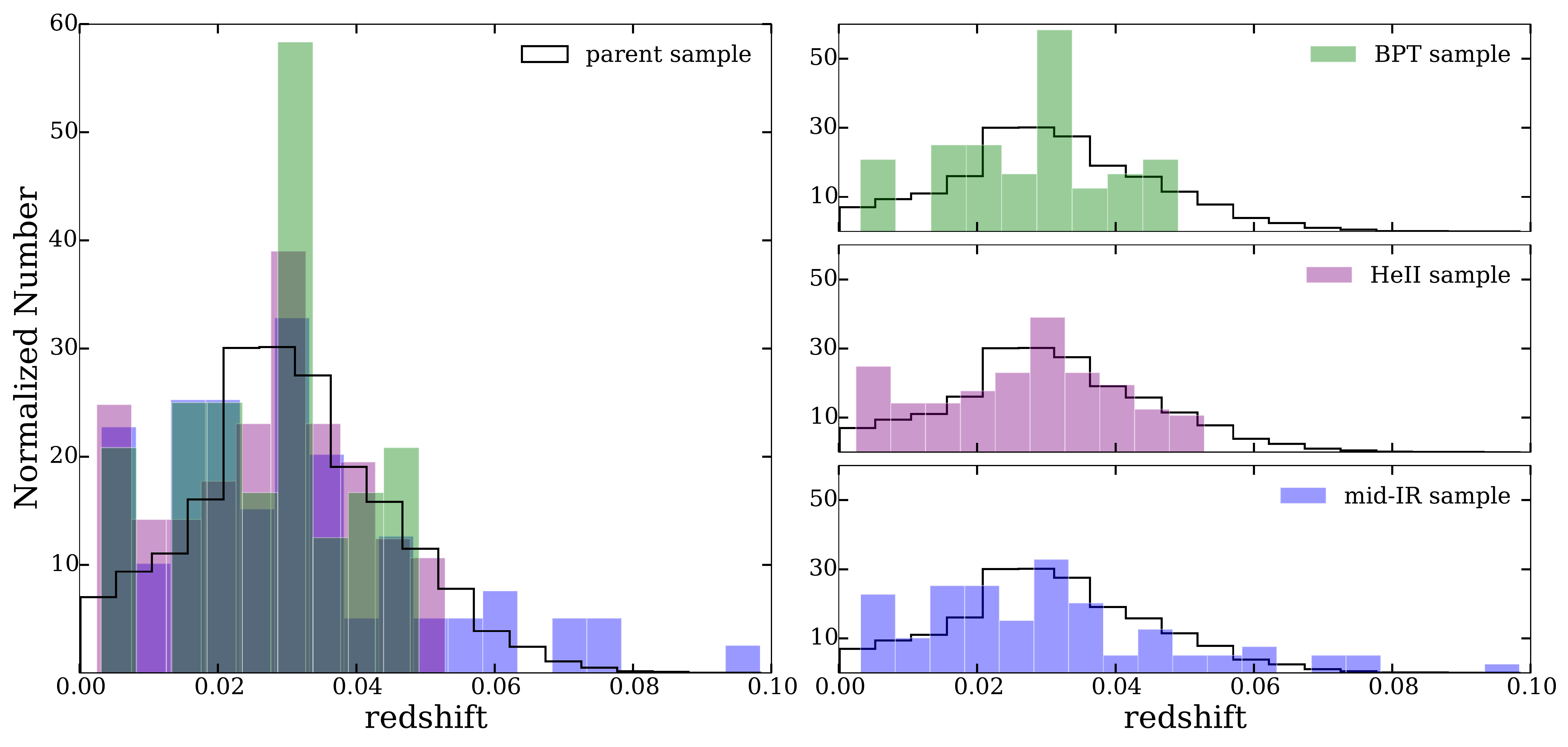}
\end{minipage}

\begin{minipage}{\textwidth}
\centering
\includegraphics[scale=0.3]{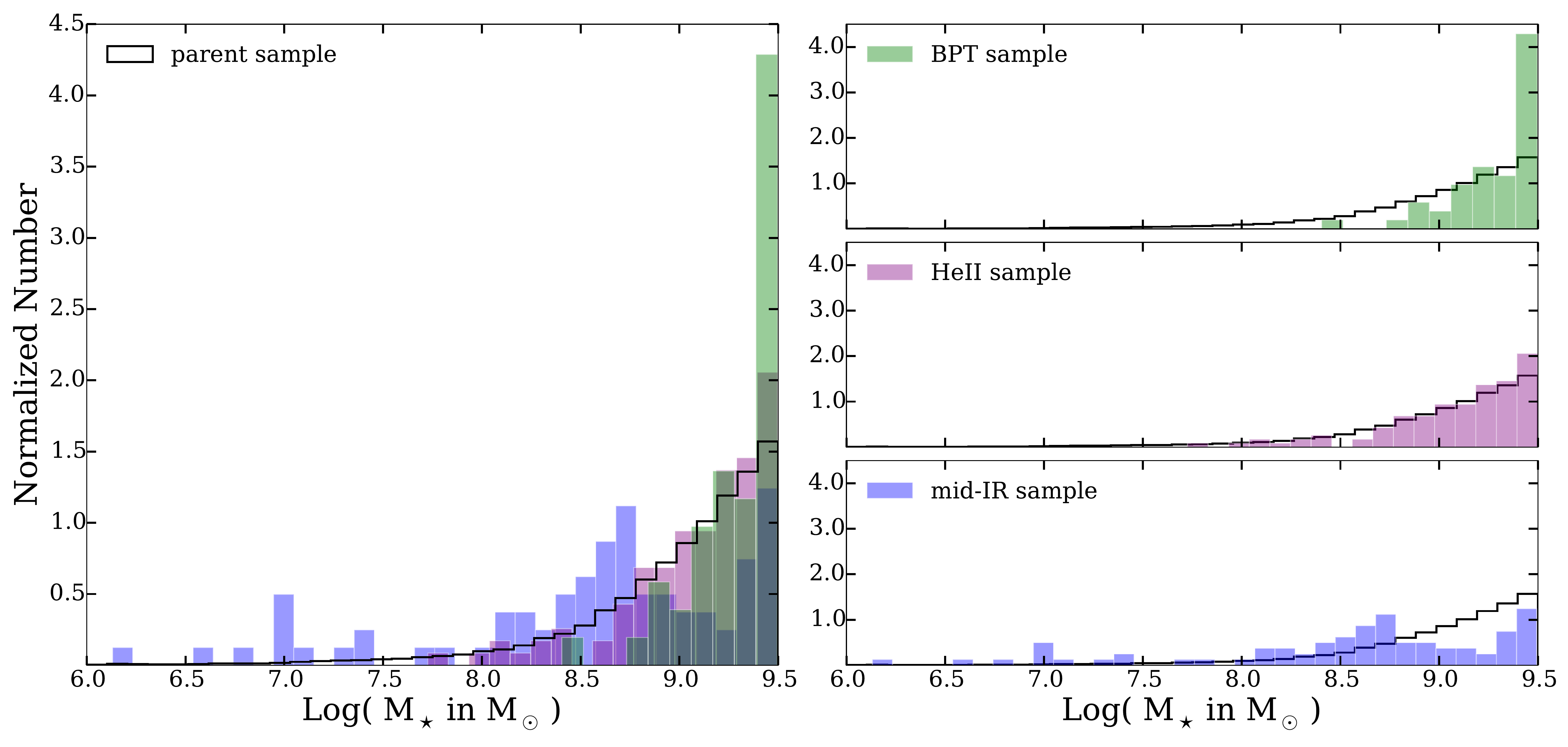}
\end{minipage}

\caption{Redshift (top panels) and stellar mass (bottom panels) distributions for the three samples of AGN candidates. The BPT sample is shown in green, the $\heii{}$ sample in purple and the mid-IR sample in blue. The black solid line corresponds to the parent sample. The histograms are normalized so that the integral over the bars is unity. Only objects with absolute $r$-band magnitude $r < 17.77$ are considered.}
\label{fig:host_1}

\end{figure*}

\begin{figure*}

\begin{minipage}{\textwidth}
\centering
\includegraphics[scale=0.445]{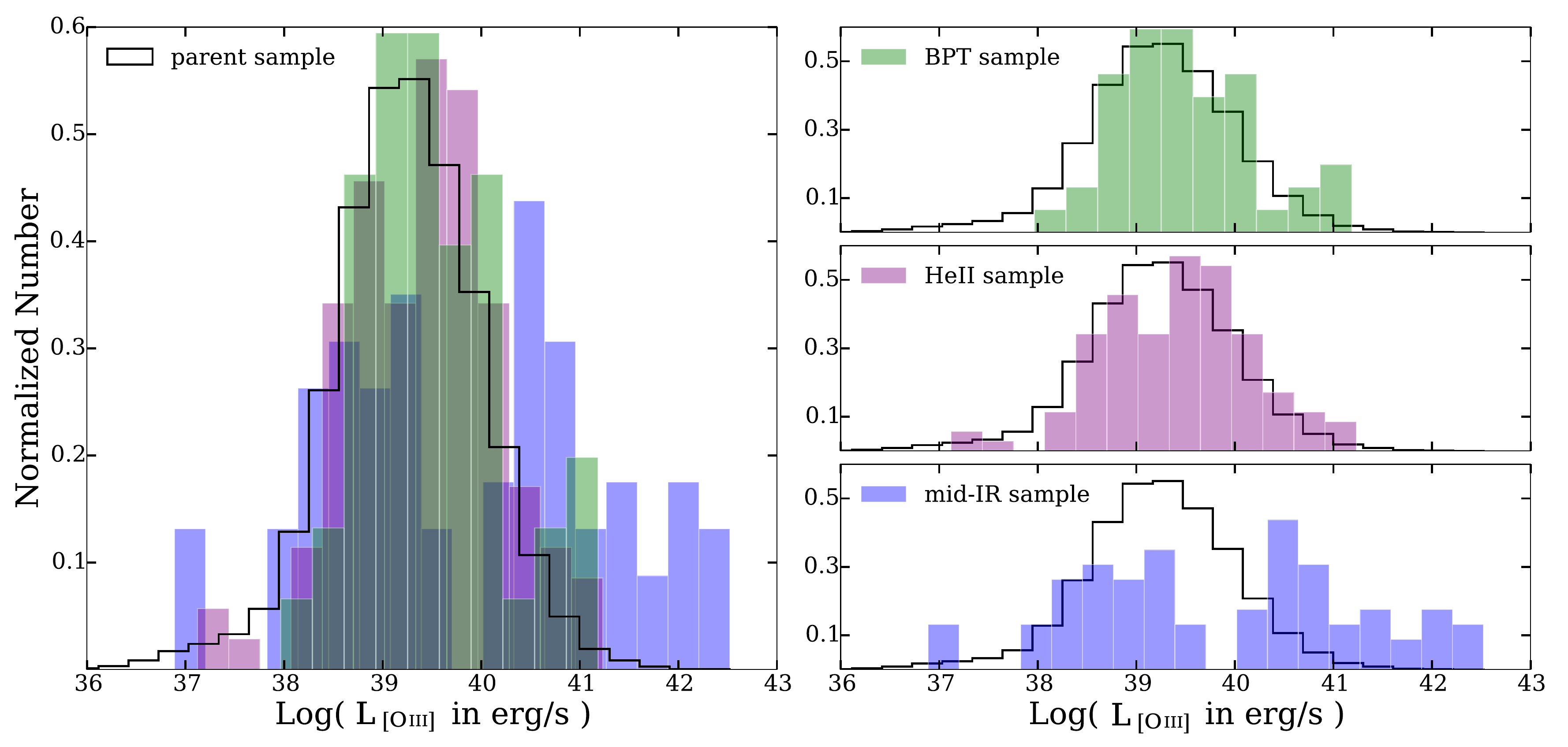}
\end{minipage}

\begin{minipage}{\textwidth}
\centering
\includegraphics[scale=0.3]{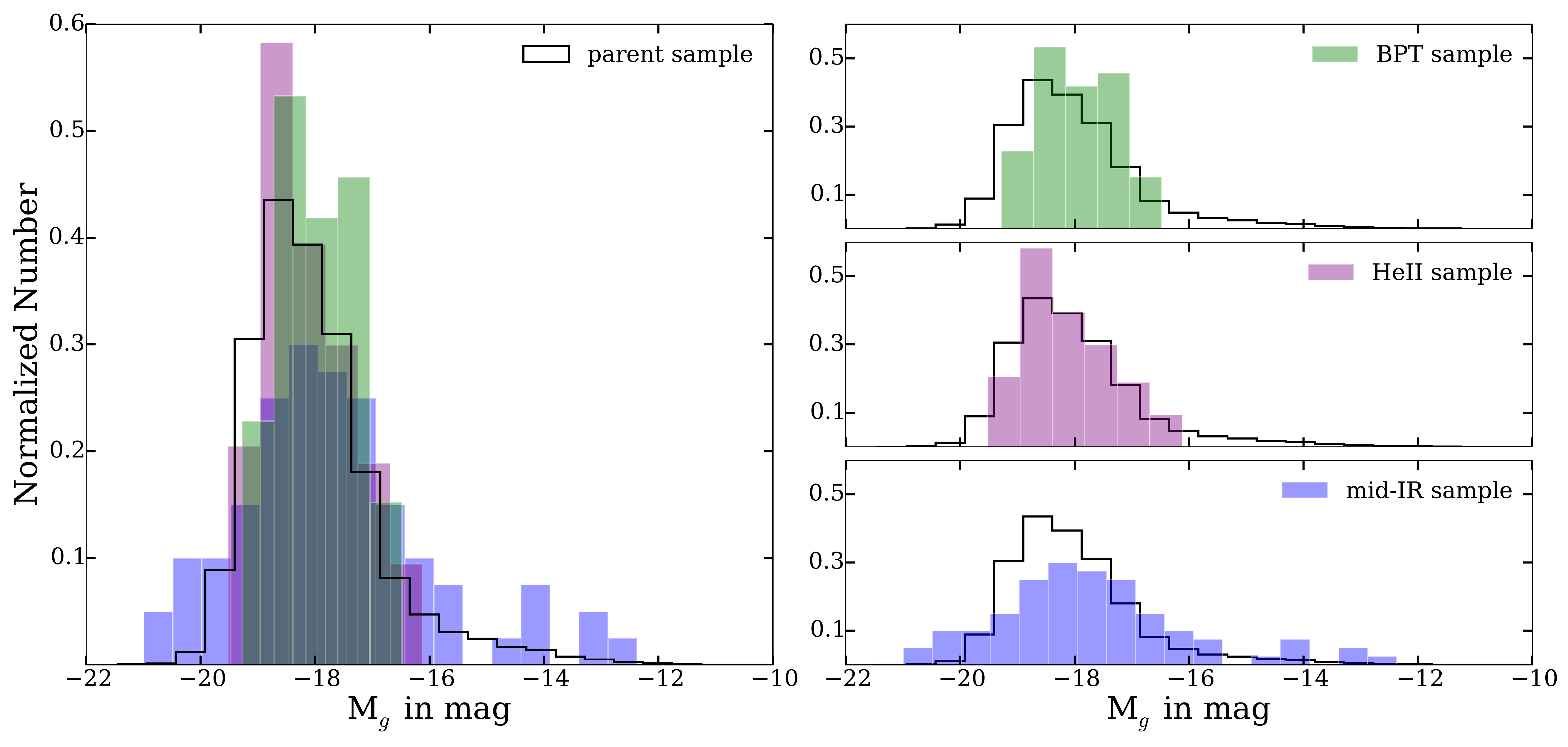}
\end{minipage}

\caption{Same as Fig. \ref{fig:host_1} but for [$\oiii{}$]$\lambda$5007 luminosity (top panels) and absolute $g$-band magnitude (bottom panels). \textcolor{black}{For the [$\oiii{}$]$\lambda$5007 distribution we considered only objects with AON∼$>$∼3 for that line. For the absolute $g$-band magnitude distribution we considered the petrosian magnitude which we corrected for galactic foreground extinction.}}
\label{fig:host_2}

\end{figure*}

\begin{figure*}

\begin{minipage}{\textwidth}
\centering
\includegraphics[scale=0.3]{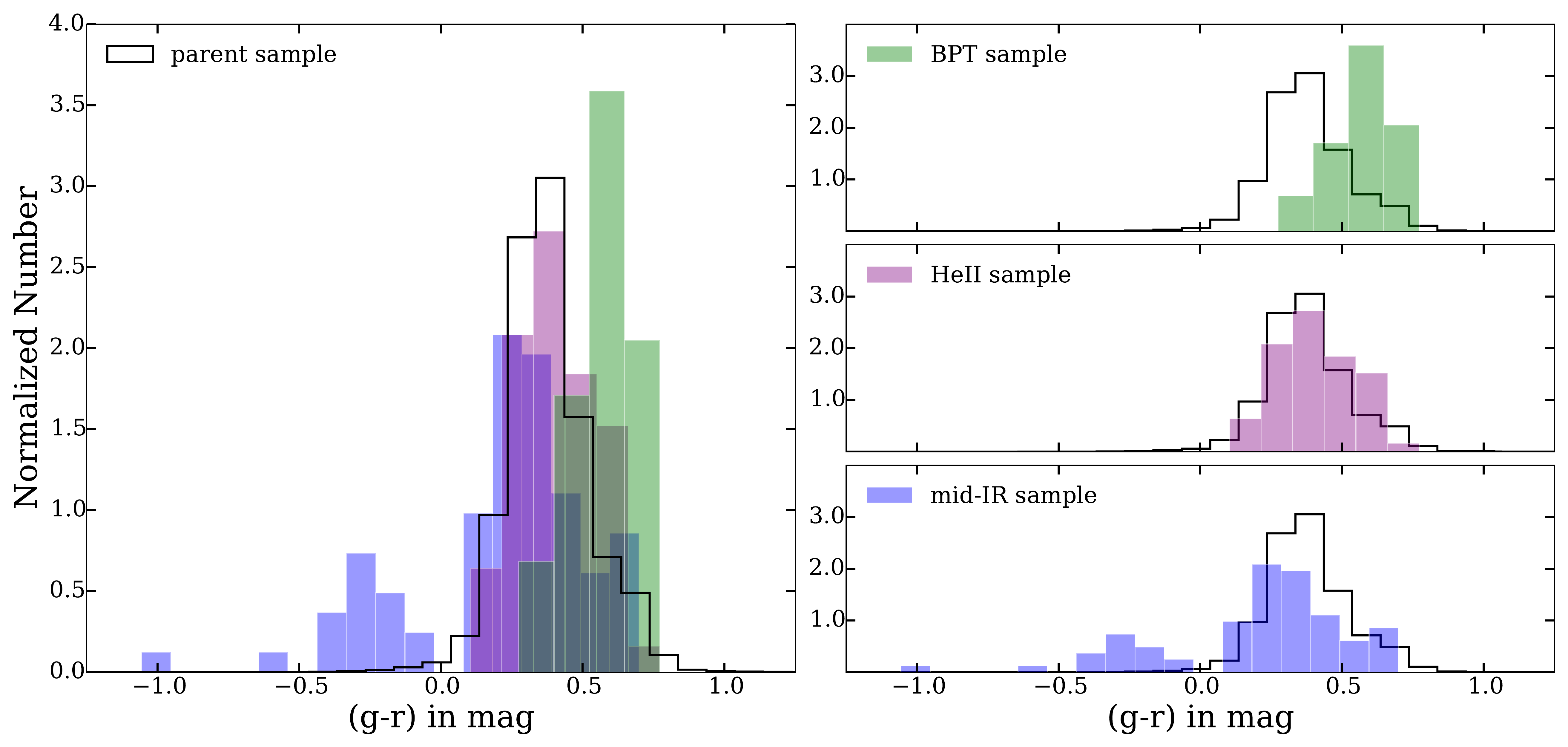}
\end{minipage}

\caption{Same as Fig. \ref{fig:host_1} but for the optical $(g-r)$ color. \textcolor{black}{We considered the petrosian $g$- and $r$-band magnitudes which we corrected for galactic foreground extinction.}}
\label{fig:host_3}

\end{figure*}

\begin{table*}
 \centering
 \begin{minipage}{140mm}
  \begin{tabular}{llcccc}
  \hline
     & & BPT sample & $\heii{}$ sample & mid-IR sample & parent sample \\
 \hline
  Redshift & $\langle z \rangle$ & 0.028 & 0.027 & 0.030 & 0.031 \\
  Stellar mass in $M_\odot$ & $\langle log(M_{\rm \star}) \rangle$ & 9.28 & 9.09 & 8.55 & 8.97 \\
  $[$OIII$]$ luminosity in erg/s & $\langle log(L_{\rm[OIII]}) \rangle$ & 39.53 & 39.43 & 39.85 & 39.23 \\
  $g$-band magnitude in mag & $\langle M_{\rm g} \rangle$ & -17.90 & -18.13 & -17.68 & -18.01 \\
  (g-r) color in mag & $\langle (g-r) \rangle$ & 0.57 & 0.41 & 0.21 & 0.39 \\
\hline
\end{tabular}
\caption{Mean values of the host galaxy properties showed in Fig. \ref{fig:host_1}, \ref{fig:host_2} and \ref{fig:host_3}. Values are given for the three AGN samples and for the parent sample separately. Only objects with absolute $r$-band magnitude $r < 17.77$ are considered.}\label{tab:host}
\end{minipage}
\end{table*}

\begin{figure*}

\begin{minipage}{\textwidth}
\centering
\includegraphics[scale=0.29]{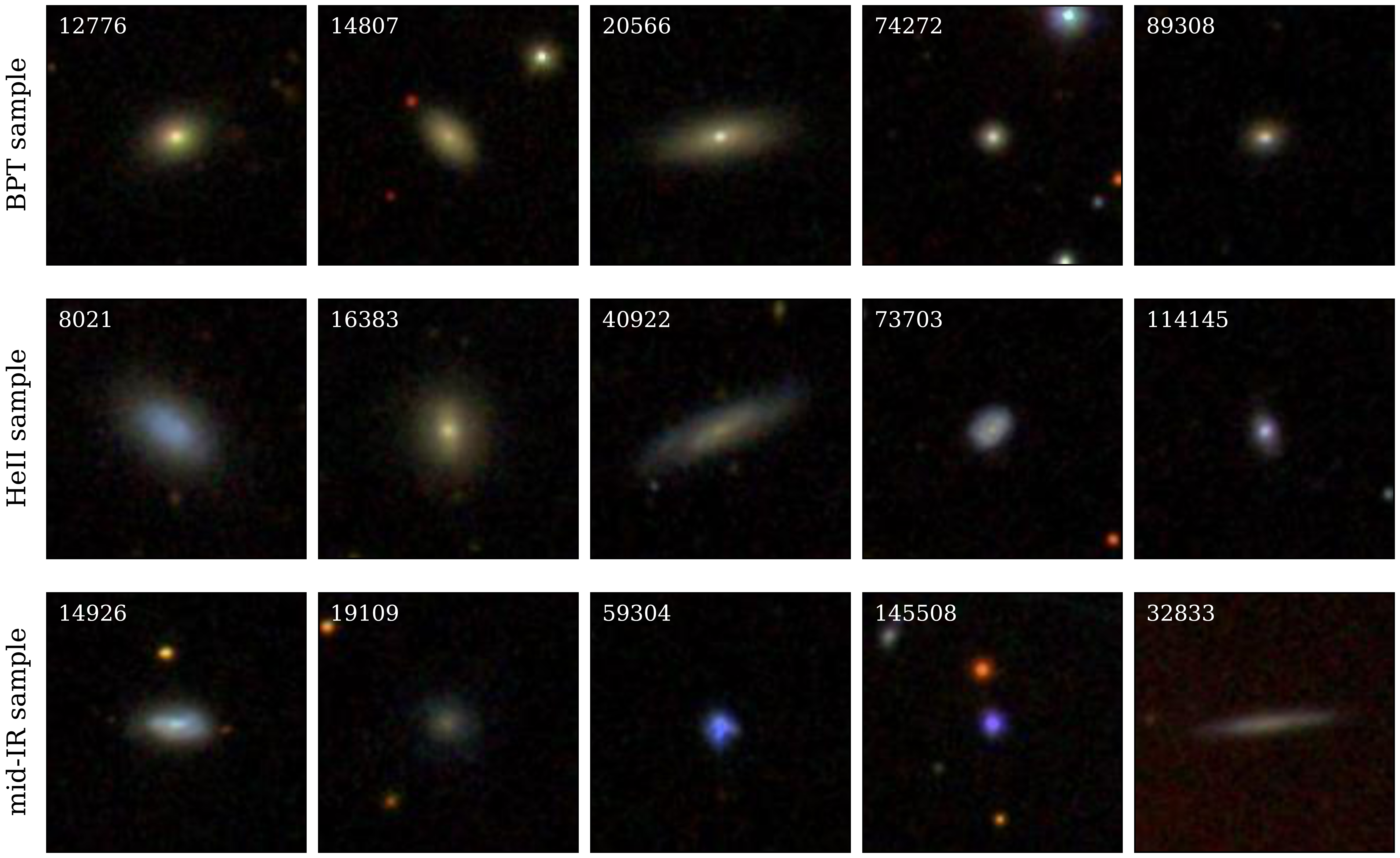}
\end{minipage}

\caption{Composite images of some of the AGN host galaxies for AGN selected with different techniques. The images are taken from {\emph{SDSS}} and the size is $50\times$50 arcseconds. \textit{Top}: BPT selection, \textit{Middle}: $\heii{}$ selection, \textit{Bottom}: mid-IR selection.}
\label{fig:host_im}

\end{figure*}







We used the galaxy parameters provided by the {\emph{SDSS}} and {\emph{MPA-JHU}} catalogs to analyse the properties of the galaxies hosting our AGN candidates.
In Fig. \ref{fig:host_1}, \ref{fig:host_2} and \ref{fig:host_3} we plotted the distributions of redshift, stellar mass, [$\oiii{}$]$\lambda$5007 luminosity, absolute $g$-band magnitude and $(g-r)$ color for the three samples (BPT selection in green, $\heii{}$ selection in purple, mid-IR selection in blue). As a comparison we also overplotted the distribution for the parent sample (black solid line). The histograms are normalized so that the integral over the bars is unity. For this analysis we considered only objects with absolute $r$-band magnitude $r < 17.77$. A summary of the results is given in Table \ref{tab:host}.\\

By sample definition, all the considered objects have redshift $z < 0.1$ and stellar mass $M_{\rm \star}$ $\leq$ $10^{9.5}M_\odot$ (Fig. \ref{fig:host_1}). Although we did not put any constraints on a minimum mass, all the objects have stellar mass higher than $M_{\rm \star}$∼$=$∼$10^{6}M_\odot$. This lower limit is a consequence of the {\emph{SDSS}} spectroscopic flux limit. The redshift distribution is similar for the mid-IR selected AGN and the parent sample, while BPT and $\heii{}$ selected AGN are found only at redshift $z < \textcolor{black} {0.05}$. BPT and $\heii{}$ selected AGN are found at higher masses compared to the mid-IR sample.

The [$\oiii{}$]$\lambda$5007 luminosity distribution (Fig. \ref{fig:host_2}, top panel) is similar for the BPT and HeII AGN candidates, while mid-IR AGN candidates appear to have also higher luminosity. For this plot we considered only objects with AON∼$>$∼3 for the [$\oiii{}$]$\lambda$5007 line.

The mean absolute $g$-band magnitude (Fig. \ref{fig:host_2}, bottom panel) is similar for the three samples of AGN candidates and for the total sample. For this analysis we considered the petrosian magnitude, corrected for galactic foreground extinction.

Fig. \ref{fig:host_3} shows the $(g-r)$ color distribution. Also in this case the {\emph{SDSS}} magnitudes are corrected for galactic foreground extinction. BPT selected AGN show redder optical colors compared to the parent sample, while the mid-IR selected AGN are bluer. Similar to the BPT selected AGN candidates, $\heii{}$ selected AGN candidates show redder colors compared to the mid-IR selected ones, but they are still bluer than the BPT selected AGN candidates. The difference in color is also noticed in the {\emph{SDSS}} images shown in Fig. \ref{fig:host_im}.\\

In the previous analysis we looked at each parameter separately. However, most of these parameters are correlated with each other, and a simple histogram comparison can be misleading. For this reason we plotted the 2D relations of some of these quantities (see Appendix \ref{app:host}). In all the cases the BPT and $\heii{}$ selected AGN seem to separate from most of the mid-IR sample on the 2D planes (stellar mass versus $(g-r)$, [$\oiii{}$] luminosity and $g$-band magnitude, see Figures \ref{fig:mass_colors} -- \ref{fig:mass_mag}).
This suggests that mid-IR and optical selection criteria select two different classes of AGN host galaxies.

\section{X-ray and radio analysis}

Optical emission line diagnostic and mid-IR color selection provided a sample of 336 AGN candidates. However, high resolution follow-up observations at other wavelengths are required for the confirmation of AGN activity in these objects. We searched the {\emph{Chandra}} Data Archive\footnote{http://cda.harvard.edu/chaser/} and the {\emph{VLA FIRST}} Survey Catalog\footnote{http://sundog.stsci.edu} \citep{Becker1994} to get X-ray and radio observations of our sources.\\ 

We obtained archival {\emph{Chandra}} X-ray data for 11 objects in our sample (objects 5426, 94970, 166401, 230467, 306717, 190553, 338999, 12974, 54437, 5424 and 69465). Only one source, object 69465, is detected with enough counts to extract the spectrum and fit it. This source shows a soft spectrum, which can be best fitted with the {\emph{Xspec}} model {\emph{diskbb}} with temperature kT $\sim 0.35$ keV.

For the other sources with faint detections we computed the hardness ratio (HR) from the counts in the soft ($0.5 - 2.0$ keV) and hard ($2.0 - 8.0$ keV) bands. Apart from object 5424, which shows an HR of 0.6, all the other sources are soft with HR smaller than $-0.1$. For these faint sources we estimated the flux in the $0.5 - 2.0$ keV range using {\emph{WebPIMMS}}, which we used to convert the measured $0.5 - 2.0$ keV {\emph{Chandra}} count rate into flux. We repeated the calculations for both {\emph{diskbb}} and {\emph{powerlaw} models, but the result does not change significantly.

Fig. \ref{fig:x_ray} shows the $0.5 - 2.0$ keV X-ray luminosity as a function of H$\alpha$ luminosity. The dashed line corresponds to the expected X-ray luminosity assuming that all the H$\alpha$ luminosity comes from star formation. We computed this line from the SFR -- L($0.5 - 2.0$ keV) relation given in \cite{Ranalli2003}, and the SFR -- L(H$\alpha$) relation given in \cite{Kennicutt1994}. For the four BPT and/or $\heii{}$ AGN candidates the observed X-ray luminosities exceed the luminosity expected by pure star formation by at least one order of magnitude. This is a confirmation of AGN activity. In the case of mid-IR selected AGN the observed X-ray emission could come from star formation, and we cannot rule out the possibility that these objects are indeed pure star-forming galaxies.\\

\begin{figure}
\includegraphics[scale=0.4]{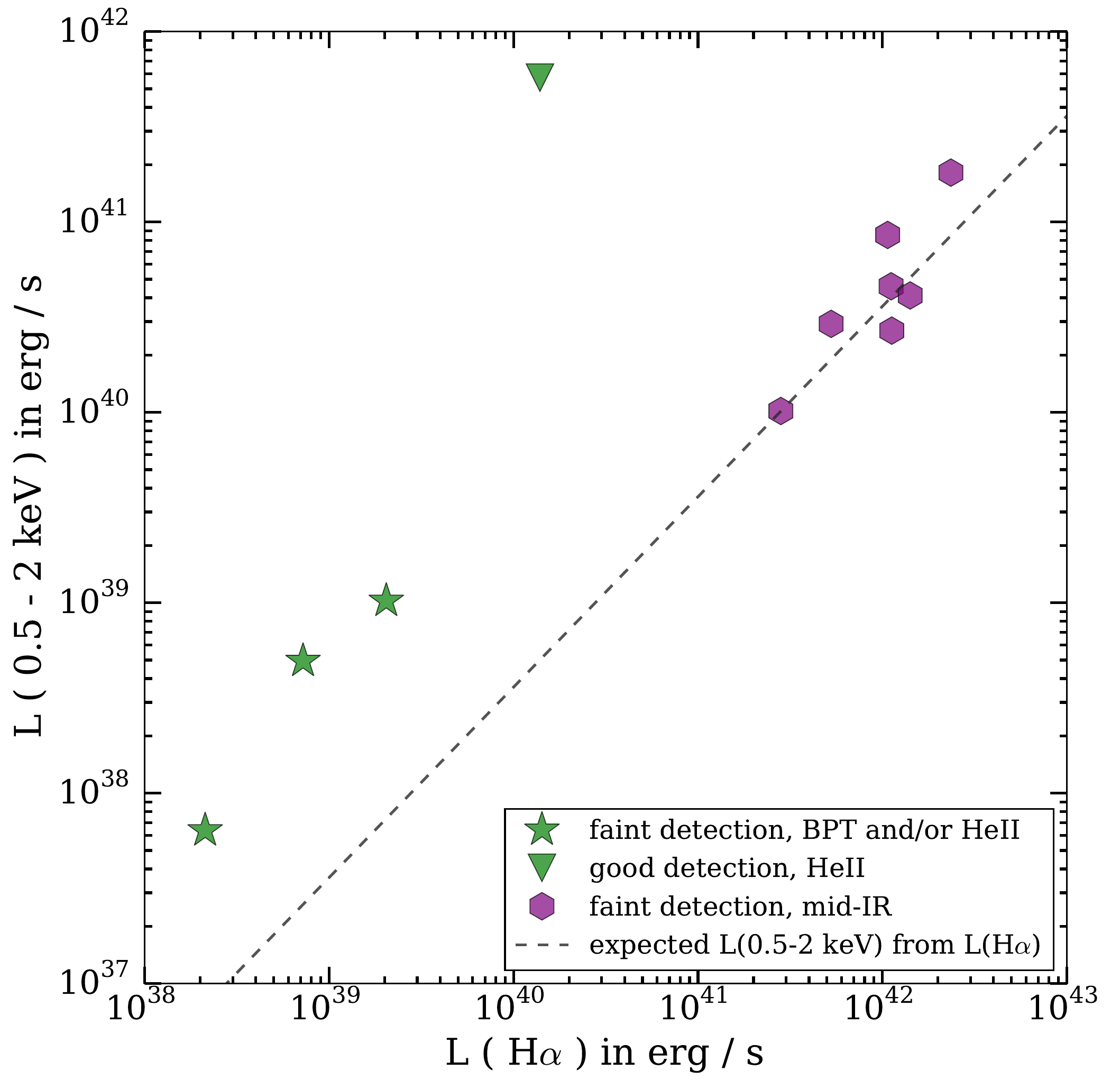}
\caption{$0.5 - 2.0$ keV X-ray luminosity as a function of H$\alpha$ luminosity. The dashed line corresponds to the expected X-ray luminosity assuming that all the H$\alpha$ luminosity comes from star formation. This line was computed from the SFR -- L($0.5 - 2.0$ keV) relation given in \protect\cite{Ranalli2003}, and the SFR -- L(H$\alpha$) relation given in \protect\cite{Kennicutt1994}. Green points correspond to BPT and/or $\heii{}$ selected AGN candidates (the stars represent the faint detections while the triagle represents the only good detection, for which it was possible to extract the X-ray spectrum), while purple hexagons represent the mid-IR selected AGN candidates.}
\label{fig:x_ray}
\end{figure}

\begin{figure}
\includegraphics[scale=0.4]{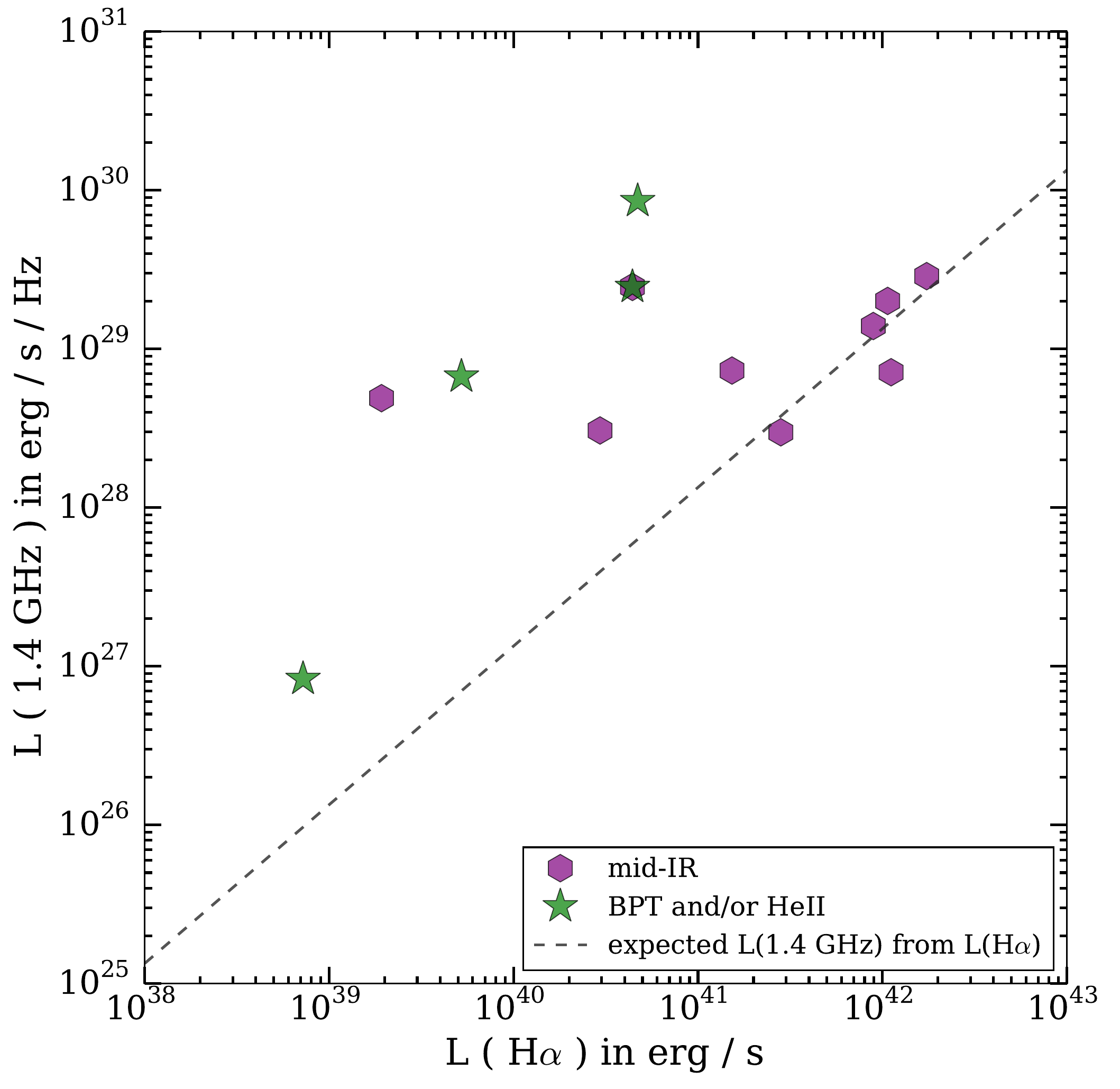}
\caption{1.4 GHz radio luminosity as a function of H$\alpha$ luminosity. The dashed line corresponds to the expected radio luminosity assuming that all the H$\alpha$ luminosity comes from star formation. This line was computed from the SFR -- L(1.4 GHz) relation given in \protect\cite{Yun2001}, and the SFR -- L(H$\alpha$) relation given in \protect\cite{Kennicutt1994}. Green stars correspond to BPT and/or $\heii{}$ selected AGN candidates, while purple hexagons represent the mid-IR selected AGN candidates.}
\label{fig:radio}
\end{figure}

The {\emph{VLA FIRST  Survey Catalog}} provided us the radio 1.4 GHz fluxes of 12 AGN candidates (5424, 84989, 92048, 54437, 60324, 72486, 94970, 115666, 190553, 221600, 50838 and 26813). Fig. \ref{fig:radio} shows the same as Fig. \ref{fig:x_ray} but for radio instead of X-ray luminosity. Here we assumed the SFR -- L(1.4 GHz) relation given in \cite{Yun2001}. Similarly to what is seen in the X-ray analysis, the radio luminosity of BPT and/or $\heii{}$ selected AGN candidates is at least one order of magnitude higher compared to what expected from stellar emission only, confirming the AGN activity. The radio luminosity seems to confirm the AGN presence also in at least three, probably four of the mid-IR AGN candidates (one of them selected also through emission line diagnostic).\\

The X-ray and radio analysis presented here are preliminary results, based on the few AGN candidates with archival X-ray and/or radio data. We are currently acquiring {\emph{XMM-Newton}} observations of some of the AGN candidates, and a more complete analysis will be presented in a following paper \textcolor{black} {(Sartori et al. in prep.)}.

\section{Discussion}\label{sec:disc}

\subsection{AGN selection and host galaxy properties} 

\textcolor{black} {In this study we selected AGN in nearby dwarf galaxies using three different selection techniques, and analysed their host galaxy properties.}


BPT selection provides \textcolor{black} {48} AGN candidates. Compared to the total sample, the galaxies hosting these AGN are found mostly at higher masses and low redshift. Moreover, they show redder optical colors. The high masses and red colors of the objects in this sample are consistent with what presented in \cite{Moran2014}. \textcolor{black} {Yet, there are a number of obstacles preventing us from detecting all the AGN in the sample using the BPT diagram. The observed trends in mass and color could therefore be attributable to selection effects and biases arising from observing strategy and detection method. The observing strategy biases are due to the fact that AGN in dwarf galaxies selected via the {\emph{SDSS}} spectra can suffer from fiber aperture effects: since dwarf galaxies are small in the sky, a big fraction of their light falls within the {\emph{SDSS}} fiber. AGN in blue star-forming low-mass dwarf galaxies may not be detected since the observed emission is dominated by galaxy light (from star formation), causing the BPT sample to be biased towards redder, more massive galaxies. Regarding the biases due to detection method, both theoretical and observational studies shows that AGN in low-metallicity galaxies, which are mainly low-mass star-forming galaxies (e.g. \citealt{Groves2006}; \citealt{Panter2008}), move to the left-hand side of the [$\nii{}$]/H$\alpha$ BPT diagram and cannot be distinguished from pure star-forming galaxies (e.g. \citealt{Groves2006}; \citealt{Kewley2013a}).}

Applying the $\heii{}$ selection diagram proposed by \cite{Shirazi2012}, we identified \textcolor{black} {121} AGN candidates. All the BPT selected candidates with well detected $\heii{}$  lines (\textcolor{black} {18}) are selected also with the Shirazi $\heii{}$ diagram. The host galaxies of the $\heii{}$ selected AGN candidates show properties similar to the BPT selected ones. In particular, they have redder colors and higher masses compared to the parent sample and to the mid-IR selected AGN sample. However, they are slightly bluer and less massive compared to the BPT selected sample, what could suggest that they are subject to a higher star formation (higher sSFR). Since [$\oiii{}$] is more affected by star formation compared to HeII, this may be the reason why most of the $\heii{}$ selected galaxies are not found in the AGN region of the BPT diagram. Thus, the Shirazi $\heii{}$ diagram appears to be more sensitive \textcolor{black} {to AGN hosted in star-forming galaxies} than the classical BPT diagram, at least in the low-mass regime. The disadvantage of this selection technique is that the $\heii{}$  line is fainter compared to the [$\oiii{}$] line, and can be well detected only in a much smaller sample.

The \textcolor{black} {189} mid-IR selected AGN candidates are found preferentially at low-masses compared to the parent sample, and show bluer optical colors. This is an indication of strong ongoing star formation. Blue colors could also signify the presence of an unobscured AGN, in which case broad H$\alpha$ lines should be visible. However, we checked the optical spectra and none of the mid-IR selected AGN candidates show broad H$\alpha$ lines. Similar to what is seen for the BPT selection, mid-IR selection is affected by important, but distinct, biases. As explained in \cite{Stern2012}, for heavily obscured low-redshift AGN or objects dominated by stellar emission the dilution by the host galaxy may cause a bluer (W1 - W2) color, so that these AGN move toward the galaxy locus and cannot be separated from star-forming galaxies. In order to test the sensitivity of mid-IR color selection we performed simulations, and found that only objects with AGN contributing at least $10-20\%$ to the total bolometric luminosity can be detected using mid-IR color-color diagram.\\

A comparison between optical emission lines (BPT and $\heii{}$) and mid-IR selection techniques shows almost no overlap (3/336 objects). Most of the BPT and $\heii{}$ selected AGN are selected as pure star-forming galaxies when applying the mid-IR color criterion. On the other hand, most of the mid-IR selected AGN candidates are located in the upper part of the star-forming region of the BPT diagram, near to the pure starburst demarcation line defined by \cite{Kewley2001}, and on the star-forming region of the Shirazi $\heii{}$ diagram. The two samples exhibit also different host galaxy properties. In summary, the mid-IR sample is both bluer and with lower stellar masses, while the optical emission lines selected sample is redder with higher stellar masses. 

One reason why mid-IR selected AGN are not selected by BPT diagram could be strong ongoing star formation.
\textcolor{black} {Although star formation could potentially affect both techniques, it is important to note that the biases due to observing strategy and detection method described for the BPT selection are not affecting the mid-IR selection. This means that star-formation may be affecting the BPT selection more than the mid-IR selection, explaining why the mid-IR selected AGN sample is bluer than the BPT selected one}. The observation that $\heii{}$ selected AGN are bluer than the BPT selected ones (but still redder than the mid-IR selected AGN) is a consequence of the fact that [$\oiii{}$] is more sensitive to star formation than HeII. Therefore, some objects can be selected as AGN using the Shirazi $\heii{}$ diagram, but the star formation is still too high to allow a selection using standard BPT diagram.

Another possible explanation of the fact that mid-IR selected AGN are bluer is that they contain less metals. Kewley et al. (2013) showed that AGN are hard to distinguish from star-forming galaxies using the BPT diagram for metallicities lower than $\log(O/H) + 12 \sim 8.4$, and $\sim 50 \%$ of the mid-IR selected AGN are below this threshold. This is suggestive that the mid-IR selection is more sensitive to detecting low-mass, low metallicity AGN host galaxies compared to optical emission lines selection.\\

\subsubsection{The mid-IR selectd AGN revisited: how many of them are AGN?}

The X-ray and radio observations available for some of the AGN candidates selected through optical emission line diagnostic confirm the AGN nature of these sources. Moreover, the BPT and $\heii{}$ regions correspondent to AGN are unlikely to be contaminated by other types of galaxies. Thus, we assume that the BPT and $\heii{}$ selected AGN candidates are indeed AGN.\\

The mid-IR selection could be contaminated by non-AGN. \cite{Stern2012} claim a realibility of $95\%$ at quasar luminosities but it is unclear to what degree this is the case for low-luminosity AGN. An alternative source of the mid-IR emission, as opposed to light reprocessed by the obscuring torus and reemitted at longer wavelenghts, is free-free emission from young stars present in extreme star-forming galaxies (e.g. \citealt{Leitherer1999}). This emission could in principle light up the dust in the galaxy in a similar way as seen in AGN. Possible contaminants are low metallicity blue compact dwarfs (e.g. \citealt{Griffith2011}), ultraluminous infrared galaxies (e.g. \citealt{Jarrett2011}), and objects similar to the Green Peas discussed in \cite{Cardamone2009}.

Four of the mid-IR selected AGN also show optical signature of BH accretion. In addition, the available radio data seem to confirm the AGN presence in at least other three objects. Thus, we know that at least some of the mid-IR selected AGN candidates are indeed AGN. However, with our analysis we cannot rule out the possibility that part of these objects are not AGN. Follow up studies, especially aiming to acquire better X-ray data for a larger part of the sample, are needed to better understand the mid-IR selection in dwarf galaxies and confirm the AGN nature of the candidates. \textcolor{black} {If some of the AGN candidates selected according to the mid-IR creteria are indeed pure star-forming galaxies, then this could further explain the bluer colors observed in these objects compared to the optically selected AGN. Moreover, it would imply a still lower AGN fraction in nearby low-mass galaxies.}



\subsection{AGN fraction in dwarf galaxies}

The measured AGN fraction depends on the selection technique. About $\textcolor{black} {0.1} \%$ of the parent sample can be selected as AGN using the BPT diagram. This value is consistent with what found by \cite{Reines2013}, where they selected 35 AGN candidates out of a sample of $\sim$25,000 objects ($\sim 0.13 \%$). On the other hand, the value is lower compared to what found by \cite{Moran2014}, but this is due to a  different selection of the parent sample. By considering also $\heii{}$ selected AGN candidates, the fraction of objects exhibiting optical signatures of BH accretion increases to $\sim \textcolor{black} {0.3} \%$. The fraction of galaxies with mid-IR colors attributable to AGN is $\sim \textcolor{black} {0.4} \%$. If all the candidates are confirmed to be AGN, this means that the mid-IR detection rate is higher than optical detection rate. For higher masses (above $\sim 10^{9.5} M_\odot$) the AGN occupation fraction determined from BPT selection is expected to be $\sim 5 - 10 \%$ (e.g. \citealt{Kauffmann2003}). Yet, including all AGN candidates found by different techniques the occupation fraction in the low-mass regime is low \textcolor{black} {measured with optical and mid-IR selection techniques}, $\sim$0.7\%. This is at least one order of magnitude below what found for more massive galaxies, \textcolor{black} {what could signify that 1) with the applied selection techniques we are missing an high number of AGN hosted in low-mass galaxies; or 2) the low-mass galaxies are less likely to host active BHs compared to more massive objects. Since AGN are active BHs, the second option could be due to a very low BH occupation fraction, which would imply also a low AGN fraction, or to a low fraction of active BHs among all the existing BHs (in this case the BH occupation fraction could still be comparable to what found at higher mass regime).}

The AGN fraction for mid-IR selected AGN shows a small decrease for increasing stellar mass, as reported also in \cite{Satyapal2014} and \cite{Marleau2014}. However, in each bin the fraction is small compared to the values found for more massive galaxies ($< \textcolor{black} {2} \%$), and the trend is not as strong as claimed by \cite{Satyapal2014}. On the other hand, the AGN fraction for optically selected AGN increases with increasing mass. \textcolor{black} {The trend showed by the optically selected AGN is consistent with what found for more massive galaxies (e.g. \citealt{Kauffmann2003}). On the other hand, the decrease seen for mid-IR selected AGN candidates could signify that in galaxies with higher stellar masses the AGN is not powerful enough to dominate over the host galaxy emission, at least in the mid-IR.} 

\subsubsection{Possible consequences for BH formation scenarios}


\textcolor{black}{Models of BH growth in a cosmological context suggest a different BH occupation fraction in the low-mass regime for different seed formation scenarios. It has therefore been proposed that measuring the BH occupation fraction in dwarf galaxies may help in discriminating between seed formation scenarios (e.g. \citealt{Volonteri2008a}; see also \citealt{Reines2013} and \citealt{Moran2014}). Since BHs in dwarf galaxies can be detected only if they are accreting (AGN), we cannot directly measure the BH occupation fraction, but we can put a lower limit on it by measuring the AGN fraction. With this study we measured a lower limit for the BH occupation fraction below 1$\%$. Despite progress, the current models are still unprecise and do not allow to estimate precisely the expected BH occupation fraction, however the difference between seed formation scenarios in general plays a role at high BH occupation fractions, which cannot be constrained by a lower limit  smaller than $1\%$. Thus, the approach of using the AGN fraction to constrain BH seed formation models, regardless of the precise predictions of the models, is unlikely to succeed.
}

\section{Summary}

We systematically searched for AGN in a sample of 48,416 nearby low-mass galaxies ($M_{\rm \star}<10^{9.5}M_\odot$, $z < 0.1$) selected from the {\emph{SDSS}} catalog. Using optical emission line diagnostic diagrams (classical BPT selection and the $\heii{}$ diagram presented in \citealt{Shirazi2012}) and mid-IR selection we assembled a sample of 336 AGN candidates. The main results of this work can be summarized as follows:

\begin{itemize}


\item The fraction of AGN in nearby dwarf galaxies selected using optical emission line diagnostic diagrams (BPT and $\heii{}$) is $\sim$0.3$\%$. For mid-IR selection the AGN fraction is $\sim$0.4$\%$. By considering all the found AGN candidates together, the observed AGN fraction in the low-mass regime seems to be at least one order of magnitude below what found for more massive objects (see e.g. \citealt{Kauffmann2003}). \textcolor{black}{Moreover, the obtained AGN fraction, which is a lower limit for the BH occupation fraction in the low-mass regime, does not allow us to put strong limits on the BH occupation fraction and to constrain currently proposed seed formation scenarios.}\\

\item Mid-IR selection and optical emission lines selection (BPT and $\heii{}$) revealed two different samples of AGN candidates. Among 336 AGN candidates, only 3 are selected by all the three criteria. The AGN candidates samples show different physical properties, what could signify also different nature. Mid-IR selected AGN are both bluer (optical color) and with lower stellar mass. This could be an indication of low metallicity and strong ongoing star formation. On the other hand, AGN selected by the BPT diagram are redder and with higher stellar mass. $\heii{}$ selected AGN candidates are bluer compared to the BPT selected candidates, but still redder than the mid-IR selected sample. \textcolor{black} {The difference in mass and color observed for the two samples could in part be a consequence of biases due to observing strategy and detection method. These biases could prevent to detect AGN in blue star-forming low-mass galaxies by applying BPT selection.}\\

\item The optical emission line diagnostic diagram based on $\heii{}$ $\lambda$4686 proposed by \cite{Shirazi2012} appears to be more sensitive \textcolor{black} {to AGN hosted in star-forming galaxies} than the classical BPT diagram, at least in the low-mass regime.\\

\item The archival X-ray and radio data available for some of the BPT and/or $\heii{}$ selected AGN candidates confirm the AGN nature of these sources. However, it is not possible to confirm that all the mid-IR selected AGN candidates are indeed AGN. A possible alternative source of the observed mid-IR emission is free-free emission from young stars present in extreme star-forming galaxies. \textcolor{black} {This could further explain the bluer colors showed by the mid-IR selected AGN candidates compared to the ones found with optical emission lines selections.} Follow-up observations, especially in the X-rays, are needed to confirm the AGN nature of these sources.\\


\item We compiled a sample of AGN candidates which will be important for future follow-up studies aiming to understand the relation between BHs and host galaxies in the low-mass regime.

\end{itemize}

This work confirmed that AGN can indeed be found in low-mass galaxies (see also e.g. \citealt{Reines2013}, \citealt{Moran2014}). Because of selection biases no selection technique based on a single wavelength range allows for a perfectly unbiased sample. Thus, only the combination of data at different wavelengths will provide us information about the nature of BHs and the relation with their host galaxies in the low-mass regime. \textcolor{black} {More} follow-up observations are needed to confirm the AGN nature of \textcolor{black} {all the} found candidates and to understand their properties.

\section*{Acknowledgments}

\textcolor{black} {We thank the referee for helpful comments, and Marta Volonteri and Michael R. Meyer for useful discussions.} LFS, KS and MK gratefully acknowledge support from Swiss National Science Foundation Professorship grant PP00P2\_138979/1 and MK support from SNSF Ambizione grant PZ00P2\_154799/1. LFS gratefully acknowledges financial support from the Swiss Study Foundation. Support for the work of ET was provided by the Center of Excellence in Astrophysics and Associated Technologies (PFB 06), by the FONDECYT regular grant 1120061 and by the CONICYT Anillo project ACT1101. KO was supported by the National Research Foundation of Korea (Doyak 2014003730). KO was a Swiss Government Excellence Scholarship holder for the academic year 2013-2014 (ESKAS No.2013.0308)\\

This research has made use of NASA's ADS service, and of data products from \emph{SDSS}, \emph{WISE}, \emph{2MASS}, \emph{Chandra} and \emph{VLA FIRST}. 
Funding for the SDSS and SDSS-II has been provided by the Alfred P. Sloan Foundation, the Participating Institutions, the National Science Foundation, the U.S. Department of Energy, the National Aeronautics and Space Administration, the Japanese Monbukagakusho, the Max Planck Society, and the Higher Education Funding Council for England. The SDSS Web Site is http://www.sdss.org/.
The SDSS is managed by the Astrophysical Research Consortium for the Participating Institutions. The Participating Institutions are the American Museum of Natural History, Astrophysical Institute Potsdam, University of Basel, University of Cambridge, Case Western Reserve University, University of Chicago, Drexel University, Fermilab, the Institute for Advanced Study, the Japan Participation Group, Johns Hopkins University, the Joint Institute for Nuclear Astrophysics, the Kavli Institute for Particle Astrophysics and Cosmology, the Korean Scientist Group, the Chinese Academy of Sciences (LAMOST), Los Alamos National Laboratory, the Max-Planck-Institute for Astronomy (MPIA), the Max-Planck-Institute for Astrophysics (MPA), New Mexico State University, Ohio State University, University of Pittsburgh, University of Portsmouth, Princeton University, the United States Naval Observatory, and the University of Washington.
We are grateful to the MPA/JHU group for access to their data products and catalogues (maintained by Jarle Brinchmann at http://www.mpa-garching.mpg.de/SDSS/).
The Wide-field Infrared Survey Explorer is a joint project of the University of California, Los Angeles, and the Jet Propulsion Laboratory/California Institute of Technology, funded by the National Aeronautics and Space Administration.
The Two Micron All Sky Survey is a joint project of the University of Massachusetts and the Infrared Processing and Analysis Center/California Institute of Technology, funded by the National Aeronautics and Space Administration and the National Science Foundation.
This research has made use of data obtained from the Chandra Data Archive and the Chandra Source Catalog, and software provided by the Chandra X-ray Center (CXC) in the application package CIAO.


\clearpage

\appendix

\section{AGN candidates}\label{app:AGN}

Tables \ref{tab:1} -- \ref{tab:2} give an overview of the AGN candidate samples. In the first column we list the object IDs used in this work. Stellar masses are given in solar masses $M_{\rm \odot}$. The [$\oiii{}$] luminosity is given in erg/s. The last four columns show by which selection criteria the sources are classified as AGN.

\setlength{\tabcolsep}{4.5pt}
\begin{table*}
 \centering
 \begin{minipage}{160mm}
  \begin{tabular}{lllllllcccc}
  \hline
    {ID} & {RA} & {DEC} & {MJD\_Plate\_Fiber} & {redshift} & {Log M$_{\rm \star}$} & {L [$\oiii{}$]} & {BPT} & {HeII} & {Mid-IR} & {Mid-IR}\\
     &  &  &  &  &  &  &  &  & {Stern} & {Jarrett}\\
 \hline
      5424$^b$$^{,c}$ & 12:07:46.11 & +43:07:34.9 & 53120\_1450\_353 & 0.0031 & 9.35 & 38.85 & $\checkmark$ & $\checkmark$ & & \\
      6868$^b$$^{,c}$ & 09:49:41.20 & +32:13:16.0 & 53432\_1946\_536 & 0.0051 & 9.23 & 39.10 & $\checkmark$ & $\checkmark$ & & \\
      7382$^b$$^{,c}$ & 14:48:42.57 & +12:27:26.0 & 53521\_1714\_376 & 0.0060 & 9.49 & 38.81 & $\checkmark$ & & & \\
      7489$^b$$^{,c}$ & 16:23:35.08 & +45:44:43.5 & 52377\_624\_111  & 0.0062 & 9.42 & 37.96 & $\checkmark$ & & & \\
      7802$^b$$^{,c}$ & 11:09:12.39 & +61:23:46.7 & 52295\_775\_309  & 0.0067 & 8.40 & 38.81 & $\checkmark$ & $\checkmark$ & & \\
      11779$^a$$^{,b}$ & 10:09:35.66 & +26:56:49.0 & 53757\_2347\_456 & 0.0143 & 8.77 & 40.20 & $\checkmark$ & $\checkmark$ & & \\
      12254$^a$ & 08:40:25.55 & +18:18:59.0 & 53711\_2278\_445 & 0.0150 & 9.14 & 39.47 & $\checkmark$ & $\checkmark$ & & \\
      12776 & 14:31:46.76 & +26:16:24.4 & 53827\_2135\_453 & 0.0156 & 9.35 & 38.81 & $\checkmark$ & & & \\
      12974$^a$$^{,b}$ & 08:11:45.30 & +23:28:25.7 & 52943\_1584\_567 & 0.0158 & 9.07 & 39.50 & $\checkmark$ & $\checkmark$ & & \\
      14523$^a$$^{,b}$$^{,c}$ & 14:05:10.40 & +11:46:16.9 & 53799\_1703\_510 & 0.0175 & 9.24 & 39.44 & $\checkmark$ & $\checkmark$ & & \\
      14807 & 00:21:45.81 & +00:33:27.3 & 51900\_390\_465  & 0.0177 & 9.16 & 38.97 & $\checkmark$ & & & \\
      16257$^a$ & 14:02:28.73 & +09:18:56.4 & 54175\_1807\_560 & 0.0192 & 8.84 & 39.29 & $\checkmark$ & & & \\
      20566$^c$ & 12:11:53.18 & +23:38:37.3 & 54210\_2644\_561 & 0.0220 & 9.50 & 40.11 & $\checkmark$ & $\checkmark$ & & \\
      22827$^a$$^{,c}$ & 13:04:57.86 & +36:26:22.2 & 53799\_2016\_414 & 0.0229 & 9.23 & 39.45 & $\checkmark$ & & & \\
      23389$^a$ & 11:43:02.41 & +26:08:19.0 & 53792\_2221\_218 & 0.0231 & 9.27 & 39.81 & $\checkmark$ & & & \\
      27247$^a$ & 11:05:03.96 & +22:41:23.5 & 54149\_2488\_561 & 0.0248 & 9.10 & 39.24 & $\checkmark$ & & & \\
      27843$^a$ & 02:48:25.27 & -00:25:41.4 & 51871\_409\_150  & 0.0251 & 8.94 & 38.95 & $\checkmark$ & & & \\
      31578$^a$ & 11:13:19.24 & +04:44:25.2 & 52326\_835\_433  & 0.0267 & 9.24 & 39.57 & $\checkmark$ & & & \\
      33647 & 09:07:37.06 & +35:28:28.4 & 52703\_1212\_97  & 0.0275 & 9.44 & 39.85 & $\checkmark$ & & & \\
      37125 & 08:42:34.51 & +03:19:30.7 & 52224\_564\_422  & 0.0288 & 9.34 & 40.21 & $\checkmark$ & $\checkmark$ & & \\
      38791 & 17:32:02.97 & +59:58:54.8 & 51792\_354\_2    & 0.0293 & 9.05 & 39.44 & $\checkmark$ & & & \\
      39324$^a$ & 08:42:04.93 & +40:39:34.5 & 52296\_829\_391  & 0.0295 & 9.20 & 39.18 & $\checkmark$ & & & \\
      40089$^a$ & 14:40:12.70 & +02:47:43.5 & 52024\_536\_575  & 0.0298 & 9.45 & 41.19 & $\checkmark$ & $\checkmark$ & $\checkmark$ & \\
      40160$^a$ & 09:02:22.76 & +14:10:49.2 & 53820\_2433\_569 & 0.0298 & 9.42 & 39.14 & $\checkmark$ & & & \\
      43170$^a$ & 14:20:44.95 & +22:42:36.9 & 54540\_2786\_552 & 0.0308 & 9.49 & 39.58 & $\checkmark$ & & & \\
      45285$^a$ & 09:21:29.99 & +21:31:39.4 & 53727\_2290\_185 & 0.0315 & 9.43 & 39.65 & $\checkmark$ & & & \\
      45874 & 16:09:43.05 & +26:47:41.8 & 52824\_1393\_505 & 0.0317 & 9.45 & 39.05 & $\checkmark$ & & & \\
      48664$^a$ & 09:54:18.16 & +47:17:25.1 & 54525\_2956\_457 & 0.0326 & 9.49 & 40.76 & $\checkmark$ & $\checkmark$ & $\checkmark$ & $\checkmark$ \\
      48794$^a$ & 14:12:08.47 & +10:29:53.9 & 53848\_1705\_98  & 0.0326 & 9.01 & 40.16 & $\checkmark$ & $\checkmark$ & & \\
      48830$^a$ & 11:44:18.84 & +33:40:07.5 & 53491\_2097\_228 & 0.0326 & 9.33 & 39.57 & $\checkmark$ & & & \\
      49040 & 01:19:05.14 & +00:37:45.1 & 51789\_398\_585  & 0.0327 & 9.26 & 39.76 & $\checkmark$ & & & \\
      49415 & 13:17:51.36 & +53:06:44.9 & 52722\_1040\_41  & 0.0328 & 9.46 & 38.90 & $\checkmark$ & & & \\
      59191 & 00:42:14.99 & -10:44:15.0 & 52162\_655\_59   & 0.0359 & 9.47 & 40.18 & $\checkmark$ & $\checkmark$ & & \\
      74272$^a$ & 13:49:39.37 & +42:02:41.4 & 52814\_1345\_41  & 0.0412 & 9.39 & 40.48 & $\checkmark$ & $\checkmark$ & & \\
      78416 & 09:44:51.02 & +12:30:44.1 & 53053\_1742\_451 & 0.0426 & 9.48 & 39.57 & $\checkmark$ & & & \\
      84989$^c$ & 13:47:36.40 & +17:34:04.6 & 54233\_2742\_442 & 0.0447 & 9.39 & 41.01 & $\checkmark$ & $\checkmark$ & & \\
      87145 & 10:44:24.58 & +18:40:09.4 & 54174\_2479\_461 & 0.0453 & 9.43 & 39.29 & $\checkmark$ & & & \\
      89308$^a$ & 15:39:41.68 & +17:14:21.9 & 54563\_2795\_509 & 0.0459 & 9.50 & 40.72 & $\checkmark$ & $\checkmark$ & & \\
      92048$^a$ & 09:06:13.77 & +56:10:15.2 & 51908\_450\_409  & 0.0466 & 9.40 & 40.89 & $\checkmark$ & $\checkmark$ & $\checkmark$ & $\checkmark$\\
      97445$^a$ & 13:04:34.92 & +7:55:05.1 & 54504\_1794\_547 & 0.0482 & 9.43 & 39.96 & $\checkmark$ & $\checkmark$ & & \\
      17389$^*$ & 11:48:47.66 & +20:10:00.5 & 54180\_2515\_357 & 0.0202 & 9.32 & 38.54 & $\checkmark$ & & & \\
      19871$^*$ & 11:42:03.18 & +16:46:34.4 & 53875\_2508\_197 & 0.0216 & 8.95 & 38.77 & $\checkmark$ & & & \\
      38107$^*$ & 03:04:51.42 & +00:12:18.2 & 51817\_411\_553  & 0.0291 & 9.29 & 39.11 & $\checkmark$ & & & \\
      59007$^*$ & 13:13:05.86 & +23:27:33.7 & 54507\_2651\_405 & 0.0359 & 9.47 & 39.24 & $\checkmark$ & & & \\
      64110$^*$ & 13:41:56.15 & +50:06:42.2 & 53433\_1669\_168 & 0.0376 & 9.11 & 38.46 & $\checkmark$ & & & \\
      73934$^*$ & 08:30:23.50 & +38:55:54.2 & 52615\_894\_20   & 0.0411 & 9.48 & 39.32 & $\checkmark$ & & & \\
      74705$^*$ & 14:32:35.92 & +14:35:21.8 & 54233\_2747\_113 & 0.0414 & 9.46 & 38.75 & $\checkmark$ & & & \\
      100261$^*$ & 02:33:46.94 & -01:01:28.3 & 51820\_407\_6    & 0.0491 & 9.49 & 40.08 & $\checkmark$ & & & \\
\hline
\end{tabular}
\caption{{\bf BPT selected AGN candidates - 1/1}. In the first column we list the object IDs used in this work. Stellar masses are given in solar masses $M_{\rm \odot}$. The [$\oiii{}$] luminosity is given in erg/s. For the objects labeled with $^*$ we assumed upper limits for H$\beta$. The last four columns show by which selection criteria the sources are classified as AGN. \newline
$^a$ Classified as AGN also in \protect\cite{Reines2013} \newline
$^b$ Classified as AGN also in \protect\cite{Moran2014} \newline
$^c$ Extended flag in WISE}
\label{tab:1}
\end{minipage}
\end{table*}

\begin{table*}
 \centering
 \begin{minipage}{160mm}
  \begin{tabular}{lllllllcccc}
  \hline
    {ID} & {RA} & {DEC} & {MJD\_Plate\_Fiber} & {redshift} & {Log M$_{\rm \star}$} & {L [$\oiii{}$]} & {BPT} & {HeII} & {Mid-IR} & {Mid-IR}\\
     &  &  &  &  &  &  &  &  & {Stern} & {Jarrett}\\
 \hline
      5041$^c$ & 01:42:26.99 & +13:58:37.1 & 51820\_429\_352  & 0.0025 & 8.01 & 37.12 &  & $\checkmark$ & & \\
      5424$^b$$^{,c}$ & 12:07:46.11 & +43:07:34.9 & 53120\_1450\_353 & 0.0031 & 9.35 & 38.85 & $\checkmark$ & $\checkmark$ & & \\
      5426$^b$$^{,c}$ & 11:51:13.43 & +50:09:24.8 & 52412\_968\_386  & 0.0031 & 9.26 & 38.30 &  & $\checkmark$ & & \\
      5634$^{c}$ & 12:08:42.33 & +36:48:10.0 & 53472\_2105\_348 & 0.0034 & 8.13 & 37.24 &  & $\checkmark$ & & \\
      6868$^b$$^{,c}$ & 09:49:41.20 & +32:13:16.0 & 53432\_1946\_536 & 0.0051 & 9.23 & 39.10 & $\checkmark$ & $\checkmark$ & & \\
      7106$^{c}$ & 14:58:46.08 & +02:58:08.6 & 52055\_589\_282  & 0.0055 & 8.21 & 38.91 &  & $\checkmark$ & & \\
      7192$^{c}$ & 10:38:01.62 & +64:15:58.9 & 51930\_489\_50   & 0.0057 & 9.13 & 39.18 &  & $\checkmark$ & & \\
      7265$^{c}$ & 16:06:41.00 & +06:34:51.4 & 53858\_1729\_120 & 0.0058 & 8.42 & 38.57 &  & $\checkmark$ & & \\
      7775$^b$$^{,c}$ & 12:25:05.61 & +05:19:44.8 & 54509\_2880\_503 & 0.0067 & 9.00 & 38.58 &  & $\checkmark$ & & \\
      7777$^{c}$ & 11:14:37.21 & +30:18:50.9 & 53460\_2092\_1   & 0.0067 & 9.41 & 38.39 &  & $\checkmark$ & & \\
      7802$^b$$^{,c}$ & 11:09:12.39 & +61:23:46.7 & 52295\_775\_309  & 0.0067 & 8.40 & 38.81 & $\checkmark$ & $\checkmark$ & & \\
      8012 & 08:26:04.39 & +45:58:03.5 & 51981\_549\_291  & 0.0071 & 9.16 & 38.58 &  & $\checkmark$ & & \\
      8021$^{c}$ & 12:53:59.23 & +29:47:19.1 & 53726\_2239\_38  & 0.0071 & 8.04 & 38.48 &  & $\checkmark$ & & \\
      8217$^{c}$ & 08:27:27.68 & +25:43:27.9 & 52945\_1586\_146 & 0.0075 & 8.62 & 38.26 &  & $\checkmark$ & & \\
      8588$^b$$^{,c}$ & 13:51:25.37 & +40:12:47.8 & 53061\_1378\_388 & 0.0082 & 9.03 & 39.96 &  & $\checkmark$ & & \\
      8807$^{c}$ & 08:12:56.37 & +54:58:08.4 & 53383\_1782\_345 & 0.0086 & 9.32 & 38.37 &  & $\checkmark$ & & \\
      9175$^{c}$ & 03:00:40.20 & +00:01:13.3 & 51817\_411\_270  & 0.0095 & 8.74 & 37.66 &  & $\checkmark$ & & \\
      9250$^{c}$ & 09:58:47.13 & +11:23:19.3 & 53055\_1744\_289 & 0.0097 & 9.04 & 38.40 &  & $\checkmark$ & & \\
      9533$^b$$^{,c}$ & 09:48:00.79 & +09:58:15.4 & 52996\_1306\_274 & 0.0104 & 8.75 & 39.28 &  & $\checkmark$ & & \\
      9690$^{c}$ & 11:18:36.35 & +63:16:50.3 & 52370\_596\_10   & 0.0107 & 9.32 & 39.56 &  & $\checkmark$ & & \\
      9874$^{c}$ & 22:10:20.63 & -09:11:31.4 & 52206\_718\_140  & 0.0110 & 8.36 & 38.51 &  & $\checkmark$ & & \\
      10219 & 00:31:27.55 & -10:40:33.2 & 52146\_654\_219  & 0.0117 & 7.73 & 39.19 &  & $\checkmark$ & & \\
      11212$^{c}$ & 13:40:36.28 & +34:17:38.2 & 53818\_2028\_312 & 0.0136 & 9.00 & 38.83 &  & $\checkmark$ & & \\
      11779$^a$$^{,b}$ & 10:09:35.66 & +26:56:49.0 & 53757\_2347\_456 & 0.0143 & 8.77 & 40.20 & $\checkmark$ & $\checkmark$ & & \\
      12254$^a$$^{,b}$ & 08:40:25.55 & +18:18:59.0 & 53711\_2278\_445 & 0.0150 & 9.14 & 39.47 & $\checkmark$ & $\checkmark$ & & \\
      12269 & 15:23:07.79 & +19:05:01.4 & 54328\_2159\_201 & 0.0150 & 8.63 & 38.97 &  & $\checkmark$ & & \\
      12840$^{c}$ & 13:19:49.95 & +52:03:41.2 & 53430\_1667\_510 & 0.0156 & 8.85 & 39.46 &  & $\checkmark$ & & \\
      12974$^a$$^{,b}$ & 08:11:45.30 & +23:28:25.7 & 52943\_1584\_567 & 0.0158 & 9.07 & 39.50 & $\checkmark$ & $\checkmark$ & & \\
      13986 & 13:57:50.41 & +26:42:40.7 & 53852\_2120\_293 & 0.0169 & 8.78 & 39.33 &  & $\checkmark$ & & \\
      14523$^a$$^{,b}$$^{,c}$ & 14:05:10.40 & +11:46:16.9 & 53799\_1703\_510 & 0.0175 & 9.24 & 39.44 & $\checkmark$ & $\checkmark$ & & \\
      15541$^c$ & 11:42:53.93 & +00:09:42.8 & 51959\_283\_426  & 0.0184 & 9.19 & 39.28 &  & $\checkmark$ & & \\
      15565 & 12:27:57.72 & +54:30:47.6 & 52707\_1019\_36  & 0.0185 & 8.30 & 38.98 &  & $\checkmark$ & & \\
      16383$^c$ & 12:03:03.59 & +56:55:00.3 & 53050\_1314\_272 & 0.0193 & 9.45 & 40.03 &  & $\checkmark$ & & \\
      17149 & 09:39:50.23 & +36:10:27.9 & 52992\_1594\_404 & 0.0200 & 8.70 & 39.53 &  & $\checkmark$ & & \\
      18265$^c$ & 12:13:45.77 & +24:21:21.8 & 54484\_2656\_21  & 0.0208 & 9.50 & 38.62 &  & $\checkmark$ & & \\
      18643$^c$ & 11:06:28.99 & +44:51:03.3 & 53054\_1436\_8   & 0.0210 & 9.48 & 39.44 &  & $\checkmark$ & & \\
      20161 & 12:06:22.54 & +31:56:29.1 & 53474\_2095\_65  & 0.0218 & 8.83 & 39.72 &  & $\checkmark$ & & \\
      20271 & 13:14:37.66 & +29:19:04.9 & 53904\_2009\_621 & 0.0218 & 9.26 & 39.48 &  & $\checkmark$ & & \\
      20566$^c$ & 12:11:53.18 & +23:38:37.3 & 54210\_2644\_561 & 0.0220 & 9.50 & 40.11 & $\checkmark$ & $\checkmark$ & & \\
      20955 & 09:04:27.03 & +29:18:56.0 & 53357\_1934\_620 & 0.0221 & 8.83 & 39.47 &  & $\checkmark$ & & \\
      22633 & 11:24:51.32 & +29:28:48.0 & 53794\_2217\_430 & 0.0228 & 9.03 & 39.55 &  & $\checkmark$ & & \\
      25147 & 11:10:23.98 & +60:51:51.8 & 52286\_774\_116  & 0.0238 & 9.17 & 39.61 &  & $\checkmark$ & & \\
      25780 & 09:44:46.54 & +13:32:23.6 & 54139\_2582\_251 & 0.0241 & 8.75 & 39.43 &  & $\checkmark$ & & \\
      26813 & 12:26:03.64 & +08:15:19.0 & 53472\_1626\_541 & 0.0246 & 9.30 & 39.57 &  & $\checkmark$ & & \\
      26925 & 12:05:50.50 & +15:54:46.4 & 53467\_1764\_527 & 0.0246 & 8.34 & 38.89 &  & $\checkmark$ & & \\
      27867 & 14:03:11.15 & +30:51:59.5 & 54180\_2121\_86  & 0.0251 & 9.15 & 38.98 &  & $\checkmark$ & & \\
      28192$^c$ & 07:55:14.82 & +53:44:43.7 & 53383\_1870\_336 & 0.0252 & 9.29 & 38.94 &  & $\checkmark$ & & \\
      28437 & 14:13:26.76 & +02:25:09.3 & 51994\_533\_396  & 0.0253 & 8.57 & 39.16 &  & $\checkmark$ & & \\
      28650$^c$ & 12:30:26.35 & +10:30:59.8 & 52731\_1232\_471 & 0.0254 & 9.42 & 38.97 &  & $\checkmark$ & & \\
      29421 & 10:48:25.08 & +38:21:10.2 & 53474\_2007\_376 & 0.0257 & 8.92 & 38.26 &  & $\checkmark$ & & \\
      29439 & 09:20:04.27 & -00:30:08.9 & 52000\_474\_313  & 0.0257 & 8.86 & 39.70 &  & $\checkmark$ & & \\
      30515 & 11:17:45.95 & +25:33:32.1 & 53794\_2214\_14  & 0.0262 & 9.26 & 38.51 &  & $\checkmark$ & & \\
      30891$^c$ & 11:46:22.21 & +49:53:21.4 & 52636\_967\_583  & 0.0264 & 9.31 & 38.78 &  & $\checkmark$ & & \\
      33190 & 11:01:47.95 & +09:53:50.4 & 52723\_1220\_559 & 0.0273 & 9.21 & 38.97 &  & $\checkmark$ & & \\
      34084$^a$ & 13:47:57.68 & +46:54:35.0 & 52736\_1284\_95  & 0.0277 & 9.36 & 40.29 &  & $\checkmark$ & & \\
      35611 & 09:21:49.44 & +23:34:38.7 & 53714\_2291\_88  & 0.0282 & 9.29 & 38.90 &  & $\checkmark$ & & \\
      36713 & 11:10:36.82 & +29:29:28.3 & 53793\_2215\_351 & 0.0286 & 8.78 & 39.79 &  & $\checkmark$ & & \\
      37026 & 12:44:33.49 & +31:05:05.8 & 53828\_2237\_607 & 0.0287 & 8.92 & 39.18 &  & $\checkmark$ & & \\
      37125 & 08:42:34.51 & +03:19:30.7 & 52224\_564\_422  & 0.0288 & 9.34 & 40.21 & $\checkmark$ & $\checkmark$ & & \\
      37135 & 15:00:40.51 & +11:56:47.4 & 53827\_1716\_474 & 0.0288 & 9.08 & 39.30 &  & $\checkmark$ & & \\
\hline
\end{tabular}
\caption{$\heii{}$ \textbf{selected AGN candidates - 1/2}. In the first column we list the object IDs used in this work. Stellar masses are given in solar masses M$_{\rm \odot}$. The [$\oiii{}$] luminosity is given in erg/s. The last four columns show by which selection criteria the sources are classified as AGN. \newline
$^a$ Classified as AGN also in \protect\cite{Reines2013}, $^b$ Classified as AGN also in \protect\cite{Moran2014} \newline
$^c$ Extended flag in WISE}
\end{minipage}
\end{table*}
\begin{table*}
 \centering
 \begin{minipage}{160mm}
  \begin{tabular}{lllllllcccc}
  \hline
    {ID} & {RA} & {DEC} & {MJD\_Plate\_Fiber} & {redshift} & {Log M$_{\rm \star}$} & {L [$\oiii{}$]} & {BPT} & {HeII} & {Mid-IR} & {Mid-IR}\\
     &  &  &  &  &  &  &  &  & {Stern} & {Jarrett}\\
 \hline
      37577 & 13:46:55.01 & +22:58:15.0 & 54230\_2666\_10  & 0.0289 & 9.26 & 38.09 &  & $\checkmark$ & & \\
      37767 & 13:31:10.68 & -01:57:01.3 & 52426\_911\_470  & 0.0290 & 9.38 & 38.52 &  & $\checkmark$ & & \\
      37935 & 10:54:30.41 & +57:22:09.7 & 52427\_949\_133  & 0.0290 & 8.94 & 39.25 &  & $\checkmark$ & & \\
      38474 & 07:55:38.19 & +24:01:03.6 & 52669\_1204\_474 & 0.0292 & 8.87 & 38.73 &  & $\checkmark$ & & \\
      39909 & 03:26:14.06 & -01:04:41.8 & 51901\_414\_240  & 0.0297 & 8.97 & 39.46 &  & $\checkmark$ & & \\
      40089$^a$ & 14:40:12.70 & +2:47:43.5 & 52024\_536\_575  & 0.0298 & 9.45 & 41.19 & $\checkmark$ & $\checkmark$ & $\checkmark$ & \\
      40922 & 10:52:40.65 & +19:56:41.3 & 54175\_2482\_404 & 0.0301 & 9.39 & 38.53 &  & $\checkmark$ & & \\
      42593 & 16:43:25.94 & +39:19:52.0 & 52395\_818\_556  & 0.0306 & 9.20 & 39.84 &  & $\checkmark$ & & \\
      43283 & 12:22:42.20 & +51:14:49.2 & 52644\_971\_325  & 0.0308 & 9.40 & 39.08 &  & $\checkmark$ & & \\
      43400 & 10:03:59.65 & +41:29:04.6 & 52672\_1217\_108 & 0.0309 & 9.28 & 39.42 &  & $\checkmark$ & & \\
      44006 & 09:15:23.86 & +6:23:02.0 & 52652\_1193\_564 & 0.0311 & 8.72 & 39.75 &  & $\checkmark$ & & \\
      45715 & 17:16:50.09 & +57:04:55.5 & 51788\_355\_160  & 0.0316 & 9.47 & 39.03 &  & $\checkmark$ & & \\
      47952 & 11:33:27.78 & +45:51:32.1 & 53050\_1442\_97  & 0.0323 & 9.02 & 39.37 &  & $\checkmark$ & & \\
      48274 & 16:20:01.08 & +08:22:58.6 & 54585\_2529\_69  & 0.0324 & 9.43 & 38.54 &  & $\checkmark$ & & \\
      48473 & 13:19:39.35 & +26:47:16.9 & 53795\_2244\_352 & 0.0325 & 9.08 & 39.68 &  & $\checkmark$ & & \\
      48664$^a$ & 09:54:18.16 & +47:17:25.1 & 54525\_2956\_457 & 0.0326 & 9.49 & 40.76 & $\checkmark$ & $\checkmark$ & $\checkmark$ & $\checkmark$\\
      48794$^a$ & 14:12:08.47 & +10:29:53.9 & 53848\_1705\_98  & 0.0326 & 9.01 & 40.16 & $\checkmark$ & $\checkmark$ & & \\
      51040 & 09:24:03.82 & +29:07:01.4 & 53389\_1939\_265 & 0.0333 & 9.48 & 38.86 &  & $\checkmark$ & & \\
      51329 & 15:12:45.01 & +26:34:49.3 & 54242\_2157\_240 & 0.0334 & 8.79 & 39.72 &  & $\checkmark$ & & \\
      51623 & 16:56:01.15 & +19:48:38.3 & 53172\_1567\_361 & 0.0335 & 8.78 & 39.95 &  & $\checkmark$ & & \\
      51844 & 11:25:54.17 & +31:53:48.4 & 53431\_1979\_575 & 0.0336 & 8.89 & 39.91 &  & $\checkmark$ & & \\
      52164 & 08:19:04.91 & +17:34:23.7 & 53726\_2271\_239 & 0.0337 & 9.12 & 39.29 &  & $\checkmark$ & & \\
      52742 & 10:10:19.10 & +29:08:05.8 & 53757\_2348\_462 & 0.0338 & 9.10 & 39.76 &  & $\checkmark$ & & \\
      53841 & 15:36:08.38 & +25:27:03.8 & 53823\_2163\_618 & 0.0342 & 9.46 & 38.86 &  & $\checkmark$ & & \\
      54785 & 14:33:46.88 & +36:26:51.2 & 53115\_1382\_305 & 0.0345 & 9.45 & 40.21 &  & $\checkmark$ & & \\
      57701 & 11:47:56.66 & +9:05:41.8 & 52734\_1226\_138 & 0.0354 & 8.89 & 39.58 &  & $\checkmark$ & & \\
      58986$^c$ & 13:10:41.07 & +44:01:07.4 & 53084\_1375\_233 & 0.0359 & 9.43 & 40.45 &  & $\checkmark$ & & \\
      59191 & 00:42:14.99 & -10:44:15.0 & 52162\_655\_59   & 0.0359 & 9.47 & 40.18 & $\checkmark$ & $\checkmark$ & & \\
      60982 & 16:16:47.19 & +18:01:09.4 & 54585\_2968\_78  & 0.0366 & 9.28 & 39.68 &  & $\checkmark$ & & \\
      63123 & 11:21:49.60 & +11:54:08.1 & 53062\_1605\_105 & 0.0373 & 9.19 & 39.68 &  & $\checkmark$ & & \\
      63673 & 13:01:43.44 & +04:08:36.7 & 52439\_849\_212  & 0.0375 & 9.22 & 39.72 &  & $\checkmark$ & & \\
      65042 & 14:04:00.01 & +10:38:31.7 & 53799\_1703\_143 & 0.0379 & 9.39 & 39.79 &  & $\checkmark$ & & \\
      65837 & 14:52:48.17 & +23:46:18.6 & 53770\_2144\_615 & 0.0382 & 9.25 & 39.71 &  & $\checkmark$ & & \\
      66137 & 15:26:37.36 & +06:59:41.7 & 54540\_1819\_496 & 0.0383 & 9.41 & 40.35 &  & $\checkmark$ & & $\checkmark$ \\
      67600 & 08:14:31.33 & +08:12:50.1 & 54055\_2571\_306 & 0.0388 & 9.49 & 39.99 &  & $\checkmark$ & & \\
      69278 & 14:48:31.55 & +16:53:28.3 & 54554\_2777\_243 & 0.0395 & 9.33 & 40.04 &  & $\checkmark$ & & \\
      69465 & 15:34:25.59 & +04:08:06.7 & 52026\_593\_547  & 0.0395 & 9.38 & 39.57 &  & $\checkmark$ & & \\
      71190 & 15:40:28.95 & +34:33:40.0 & 52872\_1402\_182 & 0.0401 & 8.69 & 39.90 &  & $\checkmark$ & & \\
      73703 & 08:51:25.82 & +39:35:41.7 & 52669\_1198\_392 & 0.0410 & 9.31 & 40.31 &  & $\checkmark$ & & \\
      74024 & 09:51:18.31 & +39:41:54.4 & 52765\_1277\_361 & 0.0411 & 9.00 & 39.95 &  & $\checkmark$ & & \\
      74272$^a$ & 13:49:39.37 & +42:02:41.4 & 52814\_1345\_41  & 0.0412 & 9.39 & 40.48 & $\checkmark$ & $\checkmark$ & & \\
      75308 & 11:49:39.99 & +36:57:28.7 & 53436\_2035\_411 & 0.0416 & 9.14 & 40.18 &  & $\checkmark$ & & \\
      78715 & 14:42:52.78 & +20:54:51.7 & 54555\_2789\_485 & 0.0427 & 8.92 & 40.56 &  & $\checkmark$ & & \\
      82176 & 10:21:17.14 & +15:05:24.2 & 54175\_2590\_206 & 0.0438 & 9.32 & 40.04 &  & $\checkmark$ & & \\
      84207 & 11:04:44.71 & +63:16:42.5 & 52370\_596\_290  & 0.0444 & 9.07 & 39.40 &  & $\checkmark$ & & \\
      84592 & 14:42:32.70 & +21:24:02.1 & 54555\_2789\_460 & 0.0445 & 9.41 & 39.62 &  & $\checkmark$ & & \\
      84989$^c$ & 13:47:36.40 & +17:34:04.6 & 54233\_2742\_442 & 0.0447 & 9.39 & 41.01 & $\checkmark$ & $\checkmark$ & & \\
      86119 & 01:56:45.30 & -00:37:37.8 & 51871\_403\_186  & 0.0450 & 9.42 & 39.91 &  & $\checkmark$ & & \\
      89308$^a$ & 15:39:41.68 & +17:14:21.9 & 54563\_2795\_509 & 0.0459 & 9.50 & 40.72 & $\checkmark$ & $\checkmark$ & & \\
      92048$^a$ & 09:06:13.77 & +56:10:15.2 & 51908\_450\_409  & 0.0466 & 9.40 & 40.89 & $\checkmark$ & $\checkmark$ & $\checkmark$ & $\checkmark$\\
      97445$^a$ & 13:04:34.92 & +07:55:05.1 & 54504\_1794\_547 & 0.0482 & 9.43 & 39.96 & $\checkmark$ & $\checkmark$ & & \\
      99464 & 16:24:35.17 & +47:32:15.8 & 52377\_624\_597  & 0.0488 & 9.12 & 40.08 &  & $\checkmark$ & & \\
      101460 & 10:41:19.81 & +14:36:41.2 & 53112\_1748\_609 & 0.0494 & 9.49 & 41.03 &  & $\checkmark$ & & \\
      101985 & 13:50:42.06 & +17:04:46.7 & 54233\_2742\_563 & 0.0496 & 9.35 & 39.67 &  & $\checkmark$ & & \\
      108738 & 01:00:05.94 & -01:10:59.0 & 51783\_395\_58   & 0.0515 & 9.47 & 40.82 &  & $\checkmark$ & & \\
      112045 & 14:03:46.96 & +33:10:10.2 & 54180\_2121\_460 & 0.0523 & 9.34 & 39.54 &  & $\checkmark$ & & \\
      114145$^a$ & 15:40:59.61 & +31:55:07.3 & 53149\_1581\_255 & 0.0528 & 9.09 & 41.23 &  & $\checkmark$ & & \\
      120037 & 09:43:45.20 & +01:30:57.5 & 51989\_480\_264  & 0.0543 & 9.39 & 39.50 &  & $\checkmark$ & & \\
      129072 & 00:58:18.14 & -08:44:15.7 & 52146\_658\_365  & 0.0569 & 9.39 & 39.67 &  & $\checkmark$ & & \\
      130041 & 11:14:28.35 & +01:44:04.3 & 52381\_510\_137  & 0.0571 & 9.43 & 39.53 &  & $\checkmark$ & & \\
      160721 & 00:24:09.49 & -01:00:32.3 & 51900\_390\_17   & 0.0645 & 9.36 & 39.54 &  & $\checkmark$ & & \\
\hline
\end{tabular}
\caption{$\heii{}$ \textbf{selected AGN candidates - 2/2}. In the first column we list the object IDs used in this work. Stellar masses are given in solar masses M$_{\rm \odot}$. The [$\oiii{}$] luminosity is given in erg/s. The last four columns show by which selection criteria the sources are classified as AGN. \newline
$^a$ Classified as AGN also in \protect\cite{Reines2013} \newline
$^c$ Extended flag in WISE}
\end{minipage}
\end{table*}


\begin{table*}
 \centering
 \begin{minipage}{160mm}
  \begin{tabular}{lllllllcccc}
  \hline
    {ID} & {RA} & {DEC} & {MJD\_Plate\_Fiber} & {redshift} & {Log M$_{\rm \star}$} & {L [$\oiii{}$]} & {BPT} & {HeII} & {Mid-IR} & {Mid-IR}\\
     &  &  &  &  &  &  &  &  & {Stern} & {Jarrett}\\
 \hline
      5449 & 12:19:23.35 & +19:35:14.4 & 54477\_2611\_266 & 0.0032 & 7.00 & - &  &  & $\checkmark$ & \\
      5619 & 12:01:22.31 & +02:11:08.3 & 52017\_516\_178  & 0.0034 & 6.13 & 39.03 &  &  & $\checkmark$ & \\
      6642 & 02:53:28.64 & -08:59:05.5 & 51901\_457\_213  & 0.0049 & 7.14 & 36.89 &  &  & $\checkmark$ & $\checkmark$\\
      6689 & 11:51:04.82 & +05:14:46.4 & 52373\_839\_623  & 0.0049 & 6.61 & 39.02 &  &  & $\checkmark$ & \\
      7289 & 15:01:00.85 & +01:00:49.9 & 52017\_539\_205  & 0.0058 & 7.35 & 36.89 &  &  & $\checkmark$ & \\
      7629 & 12:46:10.75 & +26:15:00.8 & 54205\_2238\_135 & 0.0064 & 7.41 & 39.29 &  &  & $\checkmark$ & \\
      7969 & 08:44:42.71 & +29:32:43.1 & 52937\_1269\_18  & 0.0070 & 7.29 & 37.77 &  &  & $\checkmark$ & \\
      9030 & 16:47:10.66 & +21:05:14.5 & 53149\_1570\_21  & 0.0091 & 6.76 & 40.18 &  &  & $\checkmark$ & $\checkmark$\\
      9612 & 08:42:36.58 & +10:33:13.9 & 54061\_2573\_493 & 0.0105 & 7.04 & 40.17 &  &  & $\checkmark$ & \\
      10763 & 02:18:52.90 & -09:12:18.7 & 52162\_668\_152  & 0.0128 & 7.07 & 40.27 &  &  & $\checkmark$ & $\checkmark$\\
      10897 & 11:29:25.96 & +21:21:07.6 & 54178\_2500\_180 & 0.0131 & 7.80 & 39.10 &  &  & $\checkmark$ & \\
      10957 & 10:44:57.80 & +03:53:13.1 & 52339\_578\_60   & 0.0132 & 6.96 & 40.85 &  &  & $\checkmark$ & \\
      11102 & 08:01:23.72 & +39:53:22.5 & 52201\_544\_524  & 0.0134 & 8.83 & 38.30 &  &  & $\checkmark$ & \\
      11105 & 08:22:40.29 & +03:45:46.6 & 52642\_1185\_235 & 0.0134 & 8.50 & 38.18 &  &  & $\checkmark$ & $\checkmark$\\
      11826 & 09:31:56.90 & +42:48:58.0 & 52670\_940\_332  & 0.0144 & 8.11 & 40.51 &  &  & $\checkmark$ & \\
      12643 & 08:02:22.88 & +23:24:28.9 & 52705\_1265\_306 & 0.0154 & 7.98 & 38.37 &  &  & $\checkmark$ & \\
      12868 & 13:29:32.42 & +32:34:17.1 & 53534\_2112\_35  & 0.0157 & 8.11 & 38.46 &  &  & $\checkmark$ & $\checkmark$\\
      14176 & 11:55:28.34 & +57:39:51.9 & 52790\_1313\_423 & 0.0171 & 7.73 & 41.39 &  &  & $\checkmark$ & \\
      14294 & 09:25:24.54 & +02:03:49.3 & 51965\_475\_359  & 0.0173 & 8.22 & 38.65 &  &  & $\checkmark$ & \\
      14926 & 00:59:04.10 & +01:00:04.3 & 51783\_395\_525  & 0.0178 & 8.83 & 40.35 &  &  & $\checkmark$ & $\checkmark$\\
      15020 & 12:23:16.53 & +04:50:10.1 & 54509\_2880\_230 & 0.0179 & 8.71 & 41.07 &  &  & $\checkmark$ & \\
      15156 & 11:36:51.14 & +07:40:25.2 & 53084\_1619\_549 & 0.0181 & 8.72 & 39.35 &  &  & $\checkmark$ & \\
      15175 & 08:48:20.17 & +26:24:49.9 & 53350\_1932\_188 & 0.0181 & 8.42 & 38.89 &  &  & $\checkmark$ & \\
      17350 & 09:47:22.14 & +00:44:26.9 & 51630\_266\_616  & 0.0201 & 9.31 & 39.19 &  &  & $\checkmark$ & \\
      17373 & 08:38:01.53 & +18:43:12.4 & 53705\_2277\_216 & 0.0202 & 8.58 & 37.96 &  &  & $\checkmark$ & \\
      17394 & 07:56:42.18 & +35:36:33.0 & 52238\_757\_177  & 0.0202 & 8.39 & 37.52 &  &  & $\checkmark$ & \\
      18167 & 08:24:47.50 & +18:32:43.7 & 53709\_2273\_446 & 0.0208 & 8.29 & 39.11 &  &  & $\checkmark$ & \\
      19109 & 09:15:05.13 & +28:56:41.0 & 53330\_1936\_27  & 0.0213 & 8.39 & 38.02 &  &  & $\checkmark$ & \\
      20311 & 09:16:40.98 & +18:28:07.9 & 53795\_2439\_372 & 0.0218 & 7.45 & 40.91 &  &  & $\checkmark$ & \\
      20747 & 23:53:52.51 & -00:05:55.4 & 51788\_386\_154  & 0.0220 & 9.47 & 38.12 &  &  & $\checkmark$ & \\
      20756$^c$ & 13:38:15.38 & -00:23:54.8 & 51671\_299\_222  & 0.0220 & 9.04 & 40.37 &  &  & $\checkmark$ & $\checkmark$\\
      21077 & 11:56:35.19 & +27:39:42.2 & 53819\_2226\_381 & 0.0222 & 8.32 & 38.93 &  &  & $\checkmark$ & \\
      22237 & 11:07:36.72 & +05:33:37.6 & 52356\_581\_443  & 0.0227 & 8.64 & 38.27 &  &  & $\checkmark$ & \\
      22729 & 12:10:08.35 & +44:39:06.6 & 53089\_1369\_15  & 0.0229 & 7.27 & 40.61 &  &  & $\checkmark$ & \\
      22732 & 13:28:52.96 & +15:59:34.3 & 54154\_2606\_231 & 0.0229 & 7.21 & 40.95 &  &  & $\checkmark$ & \\
      23130 & 13:41:41.61 & +35:02:14.1 & 53818\_2028\_390 & 0.0230 & 8.73 & 40.17 &  &  & $\checkmark$ & \\
      23535 & 12:10:32.56 & +44:36:04.5 & 53090\_1370\_247 & 0.0232 & 8.19 & 40.49 &  &  & $\checkmark$ & \\
      24664 & 13:12:06.27 & +5:10:59.7 & 52338\_850\_504  & 0.0236 & 8.37 & 39.14 &  &  & $\checkmark$ & \\
      24892 & 10:29:43.29 & +20:25:54.3 & 53770\_2376\_360 & 0.0237 & 8.78 & 39.41 &  &  & $\checkmark$ & \\
      25946 & 23:03:03.22 & -08:25:06.3 & 52258\_725\_447  & 0.0242 & 8.17 & 38.90 &  &  & $\checkmark$ & \\
      29484 & 22:39:31.32 & -00:40:36.2 & 52145\_377\_152  & 0.0257 & 6.96 & 38.63 &  &  & $\checkmark$ & \\
      30698 & 16:46:15.62 & +43:54:54.3 & 52051\_629\_104  & 0.0263 & 8.27 & 39.90 &  &  & $\checkmark$ & \\
      33249 & 09:14:36.22 & +18:16:02.5 & 53848\_2437\_572 & 0.0273 & 8.79 & 38.27 &  &  & $\checkmark$ & \\
      35203 & 10:54:21.65 & +2:11:36.5 & 52353\_507\_120  & 0.0281 & 8.71 & - &  &  & $\checkmark$ & \\
      35452 & 11:45:40.24 & -00:34:15.0 & 51959\_283\_164  & 0.0282 & 8.56 & 40.33 &  &  & $\checkmark$ & \\
      38058 & 07:58:24.06 & +17:07:45.7 & 53317\_1921\_638 & 0.0291 & 8.53 & 38.42 &  &  & $\checkmark$ & \\
      38410 & 10:18:57.31 & +06:25:23.1 & 52750\_998\_307  & 0.0292 & 8.77 & 37.46 &  &  & $\checkmark$ & \\
      39895 & 12:48:13.45 & +43:57:03.5 & 53063\_1373\_218 & 0.0297 & 8.78 & 40.55 &  &  & $\checkmark$ & \\
      40089$^a$ & 14:40:12.70 & +02:47:43.5 & 52024\_536\_575  & 0.0298 & 9.45 & 41.19 & $\checkmark$ & $\checkmark$ & $\checkmark$ & \\
      42470 & 16:01:35.96 & +31:13:53.8 & 52826\_1405\_395 & 0.0306 & 8.58 & 40.73 &  &  & $\checkmark$ & $\checkmark$\\
      43060 & 08:28:58.96 & +08:48:31.8 & 53084\_1758\_452 & 0.0308 & 8.56 & 39.40 &  &  & $\checkmark$ & $\checkmark$\\
      44058 & 08:45:27.61 & +53:08:52.9 & 51877\_447\_361  & 0.0311 & 8.14 & 41.43 &  &  & $\checkmark$ & \\
      45052 & 15:47:48.99 & +22:03:03.2 & 53556\_2169\_41  & 0.0314 & 7.80 & 41.26 &  &  & $\checkmark$ & \\
      48098 & 10:56:08.75 & +24:09:26.0 & 53876\_2486\_363 & 0.0324 & 8.94 & 39.22 &  &  & $\checkmark$ & \\
      48664$^a$ & 09:54:18.16 & +47:17:25.1 & 54525\_2956\_457 & 0.0326 & 9.49 & 40.76 & $\checkmark$ & $\checkmark$ & $\checkmark$ & $\checkmark$\\
      48849 & 15:09:34.17 & +37:31:46.1 & 53172\_1399\_299 & 0.0326 & 7.86 & 41.66 &  &  & $\checkmark$ & \\
      49838 & 14:23:42.87 & +22:57:28.8 & 53493\_2132\_241 & 0.0329 & 7.72 & 41.35 &  &  & $\checkmark$ & \\
      50820 & 11:32:57.32 & +31:28:31.9 & 53430\_1974\_155 & 0.0333 & 7.59 & 40.93 &  &  & $\checkmark$ & \\
      52246 & 09:58:50.00 & +67:27:08.8 & 54478\_1879\_88  & 0.0337 & 9.44 & 40.34 &  &  & $\checkmark$ & \\
\hline
\end{tabular}
\caption{\textbf{Mid-IR selected AGN candidates - 1/4}ss. In the first column we list the object IDs used in this work. Stellar masses are given in solar masses $M_{\rm \odot}$. The [$\oiii{}$] luminosity is given in erg/s. The last four columns show by which selection criteria the sources are classified as AGN. \newline
$^a$ Classified as AGN also in \protect\cite{Reines2013} \newline
$^c$ Extended flag in WISE}
\end{minipage}
\end{table*}
\begin{table*}
 \centering
 \begin{minipage}{160mm}
  \begin{tabular}{lllllllcccc}
  \hline
    {ID} & {RA} & {DEC} & {MJD\_Plate\_Fiber} & {redshift} & {Log M$_{\rm \star}$} & {L [$\oiii{}$]} & {BPT} & {HeII} & {Mid-IR} & {Mid-IR}\\
     &  &  &  &  &  &  &  &  & {Stern} & {Jarrett}\\
 \hline
      54437 & 16:21:52.57 & +15:18:56.0 & 53880\_2208\_306 & 0.0344 & 8.39 & 41.63 &  &  & $\checkmark$ & \\
      59304 & 13:03:54.44 & +37:14:01.9 & 53800\_2018\_96  & 0.0360 & 8.35 & 41.64 &  &  & $\checkmark$ & \\
      60324 & 14:16:15.62 & -01:27:52.7 & 52400\_917\_336  & 0.0363 & 8.37 & 41.23 &  &  & $\checkmark$ & \\
      64919 & 15:45:43.55 & +08:58:01.3 & 54266\_1725\_68  & 0.0379 & 8.18 & 41.96 &  &  & $\checkmark$ & \\
      66259 & 08:58:06.35 & +49:49:47.5 & 51993\_551\_76   & 0.0384 & 9.01 & 38.76 &  &  & $\checkmark$ & \\
      68371 & 09:05:31.08 & +03:35:30.4 & 52238\_566\_497  & 0.0391 & 8.08 & 41.34 &  &  & $\checkmark$ & $\checkmark$\\
      71440 & 16:19:23.61 & +25:40:10.1 & 53496\_1576\_136 & 0.0402 & 8.82 & 38.34 &  &  & $\checkmark$ & \\
      72486 & 16:39:37.27 & +27:07:50.2 & 54553\_2948\_20  & 0.0406 & 7.61 & 40.58 &  &  & $\checkmark$ & \\
      74640 & 14:18:43.41 & +28:09:57.5 & 53881\_2130\_238 & 0.0414 & 8.40 & 41.36 &  &  & $\checkmark$ & \\
      85569 & 09:40:00.52 & +20:31:22.5 & 53762\_2361\_523 & 0.0448 & 8.35 & 41.34 &  &  & $\checkmark$ & \\
      86653 & 10:20:11.34 & +01:30:46.0 & 51999\_503\_126  & 0.0451 & 9.10 & 38.71 &  &  & $\checkmark$ & \\
      86812 & 15:56:16.29 & +08:59:17.3 & 53137\_1726\_627 & 0.0452 & 9.38 & 39.60 &  &  & $\checkmark$ & $\checkmark$\\
      88704 & 10:23:52.66 & +41:52:11.0 & 53002\_1359\_412 & 0.0457 & 9.01 & 39.07 &  &  & $\checkmark$ & \\
      91423 & 15:03:10.43 & +18:43:08.8 & 54556\_2792\_284 & 0.0464 & 8.78 & 40.76 &  &  & $\checkmark$ & $\checkmark$\\
      92048$^a$ & 09:06:13.77 & +56:10:15.2 & 51908\_450\_409  & 0.0466 & 9.40 & 40.89 & $\checkmark$ & $\checkmark$ & $\checkmark$ & $\checkmark$\\
      94970 & 17:35:01.25 & +57:03:08.6 & 51818\_358\_504  & 0.0474 & 8.65 & 42.29 &  &  & $\checkmark$ & $\checkmark$\\
      95418 & 09:58:59.36 & +55:24:15.8 & 52652\_945\_423  & 0.0475 & 8.33 & 41.06 &  &  & $\checkmark$ & $\checkmark$\\
      99925 & 12:28:22.82 & +20:20:43.6 & 54477\_2611\_634 & 0.0489 & 8.08 & 41.12 &  &  & $\checkmark$ & $\checkmark$\\
      105680 & 09:02:50.25 & +14:14:10.1 & 53826\_2434\_411 & 0.0506 & 9.31 & 40.42 &  &  & $\checkmark$ & $\checkmark$\\
      105787 & 10:39:30.40 & +14:19:13.2 & 53112\_1748\_526 & 0.0507 & 8.54 & 40.76 &  &  & $\checkmark$ & $\checkmark$\\
      111723 & 13:12:53.77 & +17:12:31.2 & 54484\_2604\_439 & 0.0522 & 8.20 & 41.37 &  &  & $\checkmark$ & $\checkmark$\\
      112658 & 08:41:39.50 & +21:34:53.3 & 53360\_2084\_12  & 0.0525 & 9.37 & 39.40 &  &  & $\checkmark$ & \\
      115666 & 15:12:12.85 & +47:16:30.7 & 52721\_1050\_402 & 0.0532 & 9.15 & 41.99 &  &  & $\checkmark$ & \\
      122293 & 10:45:20.42 & +09:23:49.1 & 53491\_2147\_514 & 0.0549 & 8.90 & 41.91 &  &  & $\checkmark$ & \\
      123356 & 15:53:20.20 & +42:07:35.7 & 52764\_1334\_271 & 0.0552 & 8.60 & - &  &  & $\checkmark$ & $\checkmark$ \\
      123713 & 10:47:38.17 & +04:18:41.8 & 52339\_578\_32   & 0.0553 & 8.51 & 41.08 &  &  & $\checkmark$ & \\
      128255 & 13:35:48.08 & +37:01:45.7 & 53858\_2101\_387 & 0.0566 & 8.48 & 41.66 &  &  & $\checkmark$ & \\
      128455 & 14:15:49.61 & +01:37:56.5 & 51994\_533\_172  & 0.0567 & 8.42 & 40.87 &  &  & $\checkmark$ & $\checkmark$\\
      128605 & 01:47:07.06 & +13:56:29.1 & 51820\_429\_495  & 0.0567 & 8.45 & 41.80 &  &  & $\checkmark$ & \\
      132736 & 14:45:36.16 & +63:40:46.2 & 52339\_609\_376  & 0.0580 & 9.18 & 40.99 &  &  & $\checkmark$ & \\
      134273 & 10:32:06.02 & +22:59:21.8 & 53763\_2367\_475 & 0.0585 & 8.89 & 40.92 &  &  & $\checkmark$ & $\checkmark$\\
      137290 & 08:47:02.08 & +03:52:02.0 & 52650\_1188\_7   & 0.0593 & 9.47 & 40.67 &  &  & $\checkmark$ & \\
      137647 & 00:49:59.83 & -00:24:15.8 & 51913\_394\_269  & 0.0595 & 8.76 & 40.67 &  &  & $\checkmark$ & \\
      141121 & 13:03:03.32 & +35:51:28.6 & 53799\_2016\_394 & 0.0603 & 8.77 & 41.93 &  &  & $\checkmark$ & \\
      145508 & 10:10:42.54 & +12:55:16.8 & 53061\_1745\_475 & 0.0614 & 8.79 & 42.14 &  &  & $\checkmark$ & $\checkmark$\\
      154841 & 14:51:55.51 & +22:51:57.7 & 53770\_2144\_550 & 0.0633 & 8.31 & 41.36 &  &  & $\checkmark$ & $\checkmark$\\
      158388 & 12:39:08.59 & +09:07:17.1 & 52734\_1233\_55  & 0.0640 & 9.42 & 39.80 &  &  & $\checkmark$ & $\checkmark$\\
      160488 & 15:28:17.18 & +39:56:50.5 & 52765\_1293\_580 & 0.0645 & 8.80 & 41.78 &  &  & $\checkmark$ & \\
      164973 & 08:56:33.04 & +32:59:57.7 & 52974\_1271\_459 & 0.0654 & 9.17 & 40.38 &  &  & $\checkmark$ & \\
      164982 & 09:05:28.09 & +44:10:58.3 & 52294\_831\_526  & 0.0654 & 8.74 & 41.91 &  &  & $\checkmark$ & \\
      166401 & 14:10:59.22 & +43:02:46.9 & 53108\_1394\_294 & 0.0657 & 8.50 & 41.92 &  &  & $\checkmark$ & \\
      168543 & 23:29:36.55 & -01:10:57.0 & 51821\_384\_281  & 0.0661 & 8.81 & 41.77 &  &  & $\checkmark$ & $\checkmark$\\
      170095 & 12:36:13.64 & +09:29:03.2 & 52734\_1233\_221 & 0.0664 & 8.88 & 40.94 &  &  & $\checkmark$ & \\
      171809 & 22:25:10.13 & -00:11:52.9 & 52140\_375\_118  & 0.0668 & 8.45 & 42.00 &  &  & $\checkmark$ & \\
      176273 & 22:59:06.70 & -00:58:10.0 & 51792\_380\_253  & 0.0677 & 9.22 & 41.49 &  &  & $\checkmark$ & \\
      178309 & 12:28:08.06 & +07:54:43.4 & 53473\_1627\_426 & 0.0681 & 9.27 & 41.67 &  &  & $\checkmark$ & \\
      183496 & 10:13:34.35 & +17:52:03.1 & 54138\_2587\_565 & 0.0692 & 9.40 & 41.37 &  &  & $\checkmark$ & $\checkmark$\\
      189172 & 13:58:57.03 & +13:38:59.7 & 53883\_1778\_74  & 0.0703 & 9.35 & 40.79 &  &  & $\checkmark$ & $\checkmark$\\
      190553 & 13:44:27.36 & +56:01:29.7 & 52791\_1322\_470 & 0.0706 & 9.19 & 42.15 &  &  & $\checkmark$ & \\
      197748 & 08:48:44.51 & +51:06:26.7 & 51877\_447\_92   & 0.0719 & 8.71 & 41.35 &  &  & $\checkmark$ & \\
      199402 & 11:09:22.95 & +14:56:37.1 & 53377\_1751\_579 & 0.0721 & 9.17 & 40.73 &  &  & $\checkmark$ & $\checkmark$\\
      200366 & 08:40:00.37 & +18:05:31.0 & 53711\_2278\_411 & 0.0723 & 8.31 & 41.92 &  &  & $\checkmark$ & $\checkmark$\\
      204802 & 10:23:19.56 & +02:49:41.5 & 51999\_503\_619  & 0.0731 & 9.12 & 41.07 &  &  & $\checkmark$ & \\
      208070 & 00:06:57.02 & +00:51:26.0 & 51793\_388\_457  & 0.0737 & 8.63 & 41.89 &  &  & $\checkmark$ & \\
      208643 & 14:09:53.64 & +29:32:37.4 & 53794\_2126\_427 & 0.0738 & 9.39 & 40.63 &  &  & $\checkmark$ & $\checkmark$\\
      210237 & 11:43:48.30 & +32:42:57.9 & 53473\_2008\_505 & 0.0740 & 8.60 & 42.05 &  &  & $\checkmark$ & \\
      211304 & 07:49:47.01 & +15:40:13.3 & 53317\_1921\_281 & 0.0742 & 8.85 & 41.52 &  &  & $\checkmark$ & \\
      214163 & 09:06:28.49 & +44:58:54.5 & 52606\_898\_45   & 0.0747 & 8.83 & 41.86 &  &  & $\checkmark$ & \\
      214566 & 04:09:37.63 & -05:18:05.8 & 51910\_465\_524  & 0.0748 & 8.54 & 41.70 &  &  & $\checkmark$ & \\
\hline
\end{tabular}
\caption{\textbf{Mid-IR selected AGN candidates - 2/4}. In the first column we list the object IDs used in this work. Stellar masses are given in solar masses $M_{\rm \odot}$. The [$\oiii{}$] luminosity is given in erg/s. The last four columns show by which selection criteria the sources are classified as AGN. \newline
$^a$ Classified as AGN also in \protect\cite{Reines2013}}
\end{minipage}
\end{table*}
\begin{table*}
 \centering
 \begin{minipage}{160mm}
  \begin{tabular}{lllllllcccc}
  \hline
    {ID} & {RA} & {DEC} & {MJD\_Plate\_Fiber} & {redshift} & {Log M$_{\rm \star}$} & {L [$\oiii{}$]} & {BPT} & {HeII} & {Mid-IR} & {Mid-IR}\\
     &  &  &  &  &  &  &  &  & {Stern} & {Jarrett}\\
 \hline
      216779 & 15:18:11.87 & +19:55:14.5 & 54525\_2156\_81  & 0.0751 & 8.93 & 41.90 &  &  & $\checkmark$ & \\
      217992 & 14:55:33.68 & +00:36:57.3 & 51994\_309\_489  & 0.0753 & 9.13 & 41.41 &  &  & $\checkmark$ & $\checkmark$\\
      221600 & 08:52:21.72 & +12:16:51.7 & 53815\_2430\_117 & 0.0760 & 9.04 & 42.49 &  &  & $\checkmark$ & \\
      226725 & 14:27:22.30 & +03:25:08.1 & 52027\_585\_255  & 0.0768 & 9.12 & 41.53 &  &  & $\checkmark$ & $\checkmark$\\
      230467 & 14:09:56.77 & +54:56:48.9 & 52762\_1325\_412 & 0.0774 & 9.03 & 42.34 &  &  & $\checkmark$ & \\
      235966 & 09:28:10.52 & +15:02:27.9 & 54068\_2579\_388 & 0.0784 & 9.21 & 41.78 &  &  & $\checkmark$ & \\
      237488 & 09:36:25.35 & +05:03:32.2 & 52644\_992\_19   & 0.0786 & 8.69 & 41.84 &  &  & $\checkmark$ & \\
      240795 & 11:48:40.87 & +17:56:33.0 & 53875\_2508\_615 & 0.0792 & 8.65 & 42.22 &  &  & $\checkmark$ & \\
      245310 & 11:52:15.03 & +06:03:43.5 & 52374\_840\_446  & 0.0800 & 8.95 & 40.88 &  &  & $\checkmark$ & \\
      245739 & 00:34:13.95 & -08:58:38.5 & 52146\_654\_563  & 0.0800 & 8.82 & 40.97 &  &  & $\checkmark$ & \\
      247612 & 11:13:20.52 & +00:50:29.5 & 51984\_279\_377  & 0.0803 & 8.85 & 40.74 &  &  & $\checkmark$ & $\checkmark$\\
      247767 & 10:36:57.59 & +46:56:37.9 & 52620\_962\_227  & 0.0804 & 9.02 & 41.81 &  &  & $\checkmark$ & $\checkmark$\\
      257429 & 12:00:16.49 & +27:19:59.0 & 53819\_2226\_157 & 0.0819 & 8.59 & 42.23 &  &  & $\checkmark$ & \\
      266505 & 12:35:23.87 & +03:45:35.8 & 52381\_845\_1    & 0.0834 & 8.90 & 41.70 &  &  & $\checkmark$ & $\checkmark$\\
      272498 & 11:28:31.33 & +61:12:55.6 & 52295\_775\_29   & 0.0845 & 8.74 & 41.99 &  &  & $\checkmark$ & \\
      273004 & 10:50:32.51 & +15:38:06.3 & 53852\_2483\_254 & 0.0845 & 8.70 & 42.44 &  &  & $\checkmark$ & \\
      282992 & 14:28:05.51 & +36:27:10.4 & 53144\_1644\_564 & 0.0863 & 8.90 & 42.14 &  &  & $\checkmark$ & \\
      284918 & 02:00:51.60 & -08:45:43.0 & 52149\_666\_492  & 0.0867 & 8.97 & 41.67 &  &  & $\checkmark$ & $\checkmark$\\
      285950 & 08:25:20.11 & +08:27:23.2 & 53084\_1758\_338 & 0.0869 & 9.00 & 41.74 &  &  & $\checkmark$ & $\checkmark$\\
      296262 & 11:51:52.09 & +01:06:06.1 & 51943\_284\_448  & 0.0889 & 9.17 & 41.41 &  &  & $\checkmark$ & \\
      298133 & 09:00:47.44 & +57:42:55.2 & 51924\_483\_495  & 0.0893 & 9.24 & 41.75 &  &  & $\checkmark$ & $\checkmark$\\
      303639 & 09:26:09.41 & +12:44:49.8 & 54093\_2578\_198 & 0.0905 & 9.49 & 40.93 &  &  & $\checkmark$ & \\
      306717 & 08:44:14.22 & +02:26:21.3 & 52224\_564\_216  & 0.0912 & 9.25 & 42.34 &  &  & $\checkmark$ & \\
      309597 & 10:15:26.39 & +30:54:51.8 & 53358\_1953\_618 & 0.0918 & 9.06 & 41.95 &  &  & $\checkmark$ & $\checkmark$\\
      310250 & 14:50:56.60 & +48:37:26.5 & 52736\_1048\_424 & 0.0920 & 9.24 & 42.37 &  &  & $\checkmark$ & \\
      310490 & 08:51:15.65 & +58:40:55.0 & 54439\_1785\_201 & 0.0920 & 8.70 & 42.22 &  &  & $\checkmark$ & \\
      319086 & 09:29:18.39 & +00:28:13.2 & 52000\_474\_610  & 0.0939 & 8.63 & 41.98 &  &  & $\checkmark$ & $\checkmark$\\
      320814 & 13:29:16.55 & +17:00:20.9 & 54154\_2606\_474 & 0.0943 & 9.03 & 42.36 &  &  & $\checkmark$ & $\checkmark$\\
      325033 & 01:32:58.54 & -08:53:37.8 & 52147\_662\_466  & 0.0952 & 8.77 & 41.84 &  &  & $\checkmark$ & \\
      332236 & 16:55:14.88 & +31:29:54.3 & 52791\_1176\_198 & 0.0968 & 9.05 & 41.53 &  &  & $\checkmark$ & $\checkmark$\\
      336179 & 09:50:23.32 & +00:42:29.3 & 51608\_267\_421  & 0.0977 & 8.85 & 42.21 &  &  & $\checkmark$ & $\checkmark$\\
      337345 & 10:48:39.11 & +18:58:47.6 & 54175\_2482\_389 & 0.0980 & 9.12 & 41.89 &  &  & $\checkmark$ & \\
      338999 & 00:21:01.03 & +00:52:48.1 & 51900\_390\_445  & 0.0984 & 9.35 & 42.52 &  &  & $\checkmark$ & $\checkmark$\\
      340982 & 12:38:03.75 & +46:18:20.1 & 53089\_1455\_287 & 0.0989 & 9.09 & 42.16 &  &  & $\checkmark$ & \\
      345440 & 13:17:10.89 & +02:56:19.8 & 52295\_525\_626  & 0.0999 & 8.97 & 42.00 &  &  & $\checkmark$ & $\checkmark$\\
      5784 & 10:53:24.34 & +16:47:35.4 & 53852\_2483\_466 & 0.0036 & 7.02 & 38.11 &  &  &  & $\checkmark$ \\
      6801 & 08:52:33.75 & +13:50:28.4 & 53815\_2430\_597 & 0.0050 & 7.37 & 38.17 &  &  &  & $\checkmark$ \\
      8102 & 11:26:18.42 & +39:01:28.7 & 53436\_1996\_189 & 0.0073 & 8.07 & 37.12 &  &  &  & $\checkmark$ \\
      16073 & 00:50:28.97 & -00:31:50.0 & 51913\_394\_229  & 0.0190 & 8.28 & 39.23 &  &  &  & $\checkmark$ \\
      18330 & 12:45:07.28 & +14:19:12.1 & 53502\_1769\_40  & 0.0208 & 8.22 & 38.43 &  &  &  & $\checkmark$ \\
      20575 & 10:39:29.83 & +12:29:32.3 & 53112\_1748\_44  & 0.0220 & 8.25 & 39.18 &  &  &  & $\checkmark$ \\
      20968 & 07:56:41.82 & +28:22:18.8 & 52317\_859\_440  & 0.0221 & 8.74 & 38.50 &  &  &  & $\checkmark$ \\
      31951 & 23:01:47.70 & -09:35:04.6 & 52258\_725\_271  & 0.0268 & 8.82 & 38.15 &  &  &  & $\checkmark$ \\
      32833 & 11:08:44.78 & +36:18:33.5 & 53466\_2034\_538 & 0.0272 & 8.74 & 38.47 &  &  &  & $\checkmark$ \\
      37334 & 07:51:08.94 & +22:44:27.3 & 52669\_1203\_605 & 0.0288 & 9.27 & 39.36 &  &  &  & $\checkmark$ \\
      44317 & 16:03:01.67 & +39:37:44.3 & 52761\_1055\_403 & 0.0312 & 8.76 & 38.75 &  &  &  & $\checkmark$ \\
      44445 & 20:56:05.90 & -06:36:21.6 & 52176\_636\_176  & 0.0312 & 8.59 & 40.66 &  &  &  & $\checkmark$ \\
      45155 & 08:46:24.12 & +54:09:15.5 & 51899\_446\_154  & 0.0314 & 8.95 & 38.25 &  &  &  & $\checkmark$ \\
      46464 & 14:18:01.74 & +21:51:21.2 & 54540\_2786\_178 & 0.0319 & 8.67 & 39.36 &  &  &  & $\checkmark$ \\
      47798 & 10:28:01.15 & +08:48:51.9 & 52760\_1239\_193 & 0.0323 & 8.67 & 39.66 &  &  &  & $\checkmark$ \\
      50838 & 11:36:42.73 & +26:43:37.7 & 53816\_2219\_157 & 0.0333 & 8.94 & 38.87 &  &  &  & $\checkmark$ \\
      50950 & 10:24:29.25 & +05:24:51.0 & 52319\_575\_521  & 0.0333 & 8.22 & 41.57 &  &  &  & $\checkmark$ \\
      53767 & 15:49:27.97 & +10:47:27.0 & 54584\_2520\_166 & 0.0342 & 9.49 & 38.37 &  &  &  & $\checkmark$ \\
      57906 & 10:55:27.96 & +11:08:27.0 & 53117\_1602\_152 & 0.0355 & 8.91 & 38.69 &  &  &  & $\checkmark$ \\
      59305 & 16:21:31.69 & +10:07:30.8 & 54572\_2531\_230 & 0.0360 & 9.18 & 38.85 &  &  &  & $\checkmark$ \\
      63239 & 10:30:23.33 & +28:20:59.0 & 53794\_2353\_526 & 0.0373 & 8.57 & 40.89 &  &  &  & $\checkmark$ \\
      64962 & 12:50:39.19 & +62:10:34.8 & 52320\_782\_227  & 0.0379 & 8.63 & 39.58 &  &  &  & $\checkmark$ \\
      66137 & 15:26:37.36 & +06:59:41.7 & 54540\_1819\_496 & 0.0383 & 9.41 & 40.35 &  & $\checkmark$ &  & $\checkmark$ \\
      88228 & 08:36:35.48 & +22:17:40.5 & 53349\_1929\_74  & 0.0456 & 8.49 & 40.68 &  &  &  & $\checkmark$ \\
      90892 & 15:49:08.14 & +10:30:33.5 & 54584\_2520\_161 & 0.0463 & 9.30 & 38.64 &  &  &  & $\checkmark$ \\
\hline
\end{tabular}
\caption{{\bf{Mid-IR selected AGN candidates - 3/4}}. In the first column we list the object IDs used in this work. Stellar masses are given in solar masses $M_{\rm \odot}$. The [$\oiii{}$] luminosity is given in erg/s. The last four columns show by which selection criteria the sources are classified as AGN.}
\end{minipage}
\end{table*}
\begin{table*}
 \centering
 \begin{minipage}{160mm}
  \begin{tabular}{lllllllcccc}
  \hline
    {ID} & {RA} & {DEC} & {MJD\_Plate\_Fiber} & {redshift} & {Log M$_{\rm \star}$} & {L [$\oiii{}$]} & {BPT} & {HeII} & {Mid-IR} & {Mid-IR}\\
     &  &  &  &  &  &  &  &  & {Stern} & {Jarrett}\\
 \hline     
      94610 & 10:23:45.04 & +14:46:04.8 & 54175\_2590\_120 & 0.0473 & 9.13 & 40.67 &  &  &  & $\checkmark$ \\
      100483 & 10:35:09.33 & +09:45:16.7 & 52734\_1240\_340 & 0.0491 & 8.22 & 41.33 &  &  &  & $\checkmark$ \\
      117208 & 13:00:31.42 & +06:27:48.3 & 54504\_1794\_245 & 0.0536 & 9.36 & 38.72 &  &  &  & $\checkmark$ \\
      118838 & 11:20:40.48 & +67:46:44.0 & 51942\_491\_456  & 0.0540 & 8.48 & 40.93 &  &  &  & $\checkmark$ \\
      129926 & 13:29:50.30 & +31:11:14.2 & 53467\_2110\_466 & 0.0571 & 8.35 & 41.64 &  &  &  & $\checkmark$ \\
      137398 & 11:35:36.81 & +27:56:10.5 & 53816\_2219\_442 & 0.0594 & 9.03 & 40.36 &  &  &  & $\checkmark$ \\
      160572 & 13:06:03.91 & +19:48:22.5 & 54499\_2616\_630 & 0.0645 & 9.36 & 39.95 &  &  &  & $\checkmark$ \\
      256630 & 12:48:19.44 & +05:42:25.8 & 52426\_847\_456  & 0.0818 & 9.34 & 40.38 &  &  &  & $\checkmark$ \\
      265562 & 23:36:01.88 & -00:31:59.4 & 51821\_384\_106  & 0.0833 & 9.21 & 40.72 &  &  &  & $\checkmark$ \\
      278324 & 13:01:48.03 & +01:37:18.6 & 52027\_524\_260  & 0.0855 & 9.01 & 41.90 &  &  &  & $\checkmark$ \\
      286382 & 13:19:57.21 & +00:25:53.8 & 51984\_296\_471  & 0.0870 & 9.28 & 40.50 &  &  &  & $\checkmark$ \\ 
\hline
\end{tabular}
\caption{{\bf{Mid-IR selected AGN candidates - 4/4}}. In the first column we list the object IDs used in this work. Stellar masses are given in solar masses $M_{\rm \odot}$. The [$\oiii{}$] luminosity is given in erg/s. The last four columns show by which selection criteria the sources are classified as AGN.}
\label{tab:2}
\end{minipage}
\end{table*}

\section{Host Galaxy Properties}\label{app:host}

In Section \ref{sec:host} we analysed different host galaxy properties separately. However, most of these parameters are correlated with each other, and a simple histogram comparison can be highly misleading. For this reason Figures \ref{fig:mass_colors}, \ref{fig:mass_OIII} and \ref{fig:mass_mag} show the 2D relations of some of these quantities. In all the cases the BPT and $\heii{}$ selected AGN samples seem to separate on the 2D planes from most of the mid-IR sample. This suggests that mid-IR selection and optical selection are revealing two different classes of AGN host galaxies.

\begin{figure*}
\includegraphics[scale=0.3]{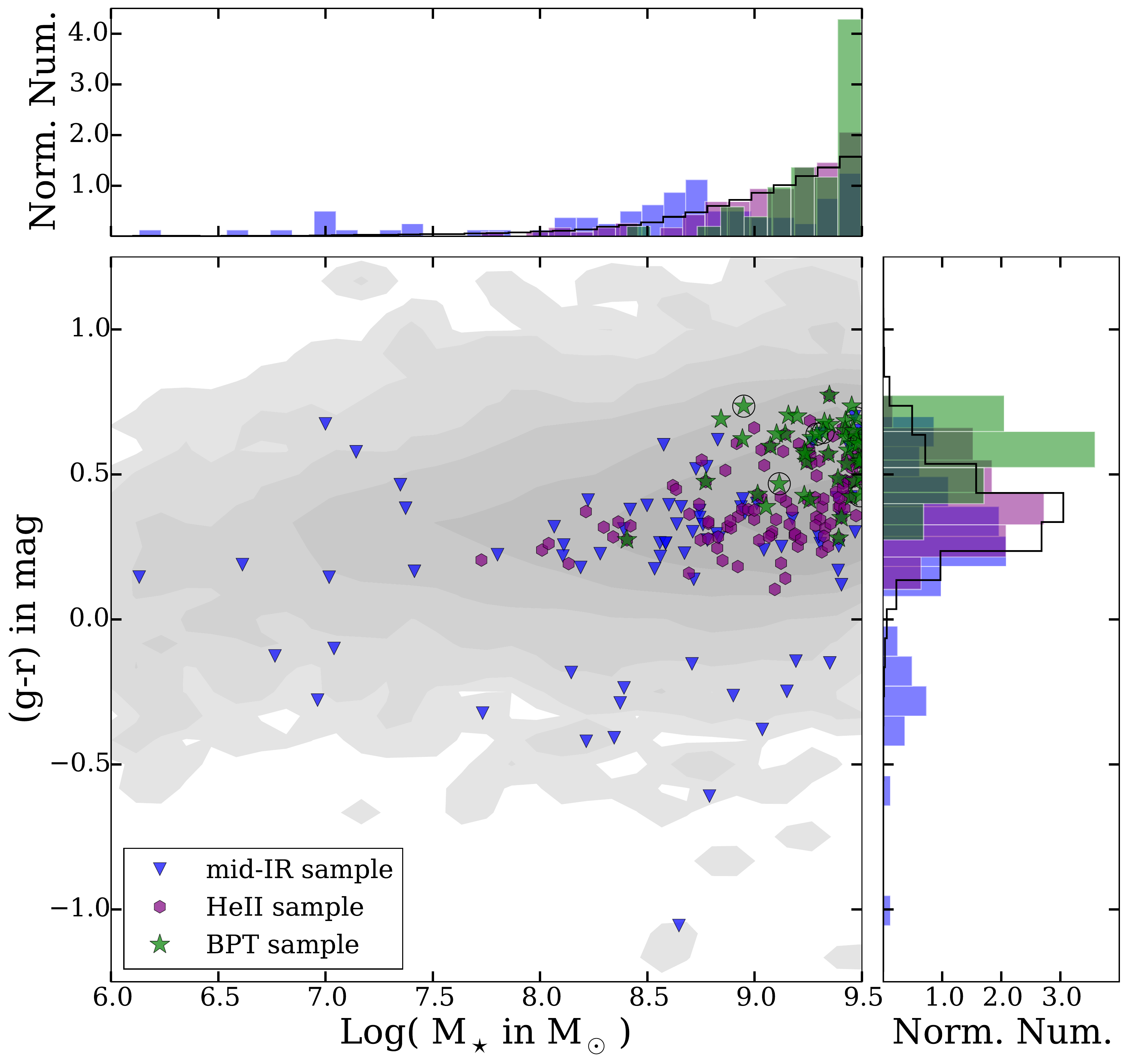}
\caption{$(g-r)$ versus stellar mass distribution for the parent sample (gray shaded area) and the AGN candidates. Green stars refer to AGN candidates selected with BPT diagram \textcolor{black}{(the circled stars show the AGN candidates with upper limits for the H$\beta$ flux)}, purple hexagons to $\heii{}$ selected AGN candidates, and blue triangles to mid-IR selected candidates. Only objects wit $r$-band magnitude $r < 17.77$ are considered. The histograms are the same as showed in Fig. \ref{fig:host_1} - \ref{fig:host_3}. The three samples of AGN candidates seem to separate on the 2D plane. In particular, mid-IR selected AGN candidates are bluer compared to the optically selected ones (BPT and $\heii{}$ selection). $\heii{}$ selected AGN candidates are bluer compared to the BPT sample, but still redder than the mid-IR sample.}
\label{fig:mass_colors}
\end{figure*}

\begin{figure*}
\includegraphics[scale=0.3]{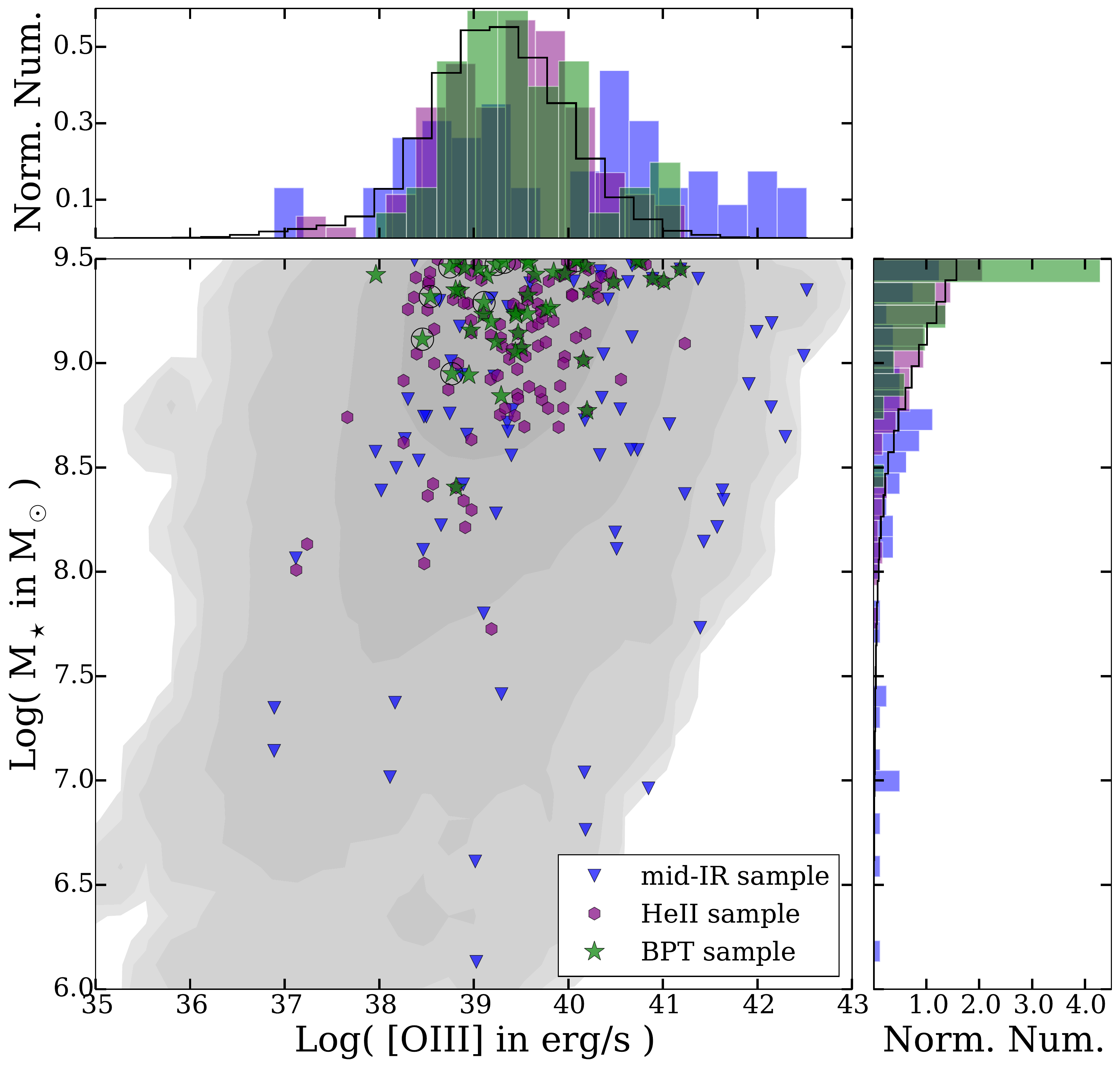}
\caption{Same as Fig. \ref{fig:mass_colors} but for the mass versus [$\oiii{}$]$\lambda$5007 distribution. Optically selected AGN candidates are found only at high masses compared to the mid-IR selected ones, and the [$\oiii{}$] luminosity is slightly lower.}
\label{fig:mass_OIII}
\end{figure*}

\begin{figure*}
\includegraphics[scale=0.3]{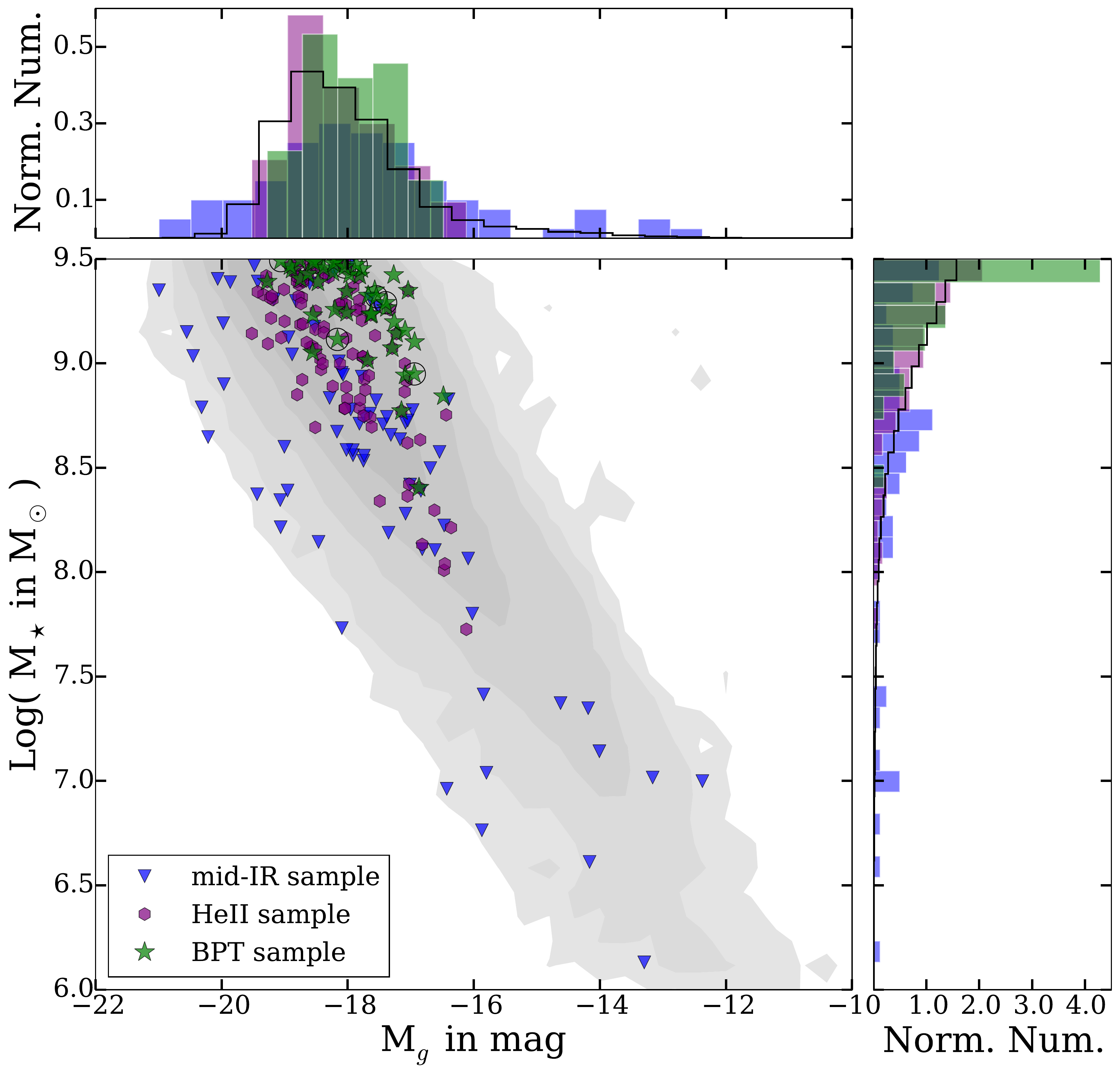}
\caption{Same as Fig. \ref{fig:mass_colors} but for the mass versus absolute $g$-band magnitude distribution. Mid-IR selected AGN candidates are found also at lower masses and at both higher and lower $g$-band magnitudes compared to the optically selected candidates.}
\label{fig:mass_mag}
\end{figure*}

\bsp

\label{lastpage}

\end{document}